\documentclass[aps,prx,reprint,superscriptaddress,longbibliography,floatfix]{revtex4-2}

\usepackage{silence}
\usepackage{graphicx}
\usepackage{dcolumn}
\usepackage{bm}
\usepackage{amsmath,amssymb}
\usepackage{amsthm}
\usepackage{physics}
\usepackage{dsfont}
\usepackage{bbm}
\usepackage[dvipsnames]{xcolor}
\usepackage{colortbl}
\usepackage[caption=false]{subfig}
\usepackage{algorithm}
\usepackage{algpseudocode}
\usepackage{algorithmicx}
\usepackage[bookmarks=false,colorlinks=true,urlcolor=RoyalBlue,citecolor=NavyBlue,linkcolor=OrangeRed]{hyperref} 
\usepackage{dsfont}
\usepackage[all]{hypcap}
\usepackage{mathtools}

\newcommand{\mC}{{\mathbf{m}_\mathcal{C}}}
\newcommand{\C}{{\mathcal{C}}}
\newcommand{\CZ}{\text{CZ}}
\newcommand{\E}[2]{\mathbb{E}_{#1}\left[ #2 \right]}
\newcommand{\Var}[2]{\text{var}_{#1}\left[ #2 \right]}
\newcommand{\ii}{{\boldsymbol{{\mathrm{i}}}}}
\newcommand{\ketp}[1]{\left|#1\right)}
\DeclarePairedDelimiterX{\kett}[1]{\lvert}{\rangle\!\rangle}{#1}

\DeclarePairedDelimiterX{\bbra}[1]{\langle\!\langle}{\rvert}{#1}
\newcommand{\kettbbra}[2]{\kett{#1}\!\bbra{#2}}

\DeclareMathOperator{\diag}{diag}

\definecolor{color1}{HTML}{648FFF}
\definecolor{color2}{HTML}{DC267F}
\definecolor{color3}{HTML}{FFB000}

\makeatletter
\long\def\@makecaption#1#2{%
  \small\raggedright
  \textbf{#1}\hspace{0.4em}\textbar\hspace{0.4em}#2\par
}
\makeatother
\usepackage[capitalise]{cleveref}
\crefname{figure}{Fig.}{Figs.}
\Crefname{figure}{Figure}{Figures}
\crefname{equation}{Eq.}{Eqs.}
\crefname{equation}{Equation}{Equations}

\newtheoremstyle{mystyle}%
{4pt}
{0pt}
{\itshape}
{}
{\bfseries}
{.}
{.2em}
{}

\theoremstyle{mystyle}
\crefname{thm}{Theorem}{Theorems}
\Crefname{thm}{Theorem}{Theorems}

\crefname{obs}{Observation}{Observations}
\Crefname{obs}{Observation}{Observations}

\crefname{lemma}{Lemma}{Lemmas}
\Crefname{lemma}{Lemma}{Lemmas}

\crefname{corll}{Corollary}{Corollaries}
\newtheorem{claim}{Claim}

\theoremstyle{remark}

\AddToHook{env/lemma/begin}{\crefalias{thm}{lemma}} 

\newcommand{\titletext}{Order from chaos with adaptive circuits on quantum hardware}
\begin{document}

\title{\titletext}
\author{Bibek Pokharel}
\email{bibek.pokharel@ibm.com}
\thanks{co-first author}
\affiliation{\vspace{-1pt}IBM Quantum, IBM Thomas J. Watson Research Center, Yorktown Heights, NY, USA}

\author{Haining Pan}
\email{haining.pan@rutgers.edu}
\thanks{co-first author}
\affiliation{Department of Physics and Astronomy, Center for Materials Theory, Rutgers University, Piscataway, NJ, USA}

\author{Kemal Aziz}
\affiliation{Department of Physics and Astronomy, Center for Materials Theory, Rutgers University, Piscataway, NJ, USA}

\author{Luke~C.~G.~Govia}
 \affiliation{\vspace{-1pt}IBM Quantum, Almaden Research Center, San Jose, CA 95126, USA}
 
\author{Sriram Ganeshan}
\affiliation{Physics Department, City College of the CUNY, New York, NY 10031}

\affiliation{CUNY Graduate Center, New York, NY 10031}

 \author{Thomas Iadecola}
\affiliation{Department of Physics and Astronomy, Iowa State University, Ames, IA 50011, USA}
\affiliation{Ames National Laboratory, Ames, IA 50011, USA}
\affiliation{Department of Physics, The Pennsylvania State University, University Park, PA 16802, USA}
\affiliation{Institute for Computational and Data Sciences, The Pennsylvania State University, University Park, PA 16802, USA}
\affiliation{Materials Research Institute, The Pennsylvania State University, University Park, PA 16802, USA}

\author{Justin H. Wilson}
\affiliation{Department of Physics and Astronomy, Louisiana State University, Baton Rouge, LA 70803, USA}
\affiliation{Center for Computation and Technology, Louisiana State University, Baton Rouge, LA 70803, USA}

\author{Barbara A. Jones}
\thanks{Deceased author. We note with sadness that Dr. Barbara Jones passed away during the preparation of this manuscript. We are grateful for her invaluable contributions to this work.}
 \affiliation{\vspace{-1pt}IBM Quantum, Almaden Research Center, San Jose, CA 95126, USA}

\author{Abhinav Deshpande}
 \affiliation{\vspace{-1pt}IBM Quantum, Almaden Research Center, San Jose, CA 95126, USA}

\author{Jedediah H. Pixley}
\email{jed.pixley@physics.rutgers.edu}
\affiliation{Department of Physics and Astronomy, Center for Materials Theory, Rutgers University, Piscataway, NJ, USA}
\affiliation{Center for Computational Quantum Physics, Flatiron Institute, NY, USA}

\author{Maika Takita}
\email{mtakita@us.ibm.com}
 \affiliation{\vspace{-1pt}IBM Quantum, IBM Thomas J. Watson Research Center, Yorktown Heights, NY, USA}

\date{\today}

\begin{abstract}
{\bf
Programmable quantum devices provide a platform to control the coherent dynamics of quantum wavefunctions. Here we experimentally realize adaptive monitored quantum circuits, which incorporate conditional feedback into non-unitary evolution, to control quantum chaotic dynamics using a combination of local mid-circuit measurements and resets. The experiments are performed with an IBM superconducting quantum processor using up to 100 qubits that samples a quantum version of the classically chaotic Bernoulli map.  This map scrambles quantum information, while local measurements and feedback attempt to steer the dynamics toward a state 
that is a fixed point of the map.  This competition drives a dynamical phase transition between quantum and classical dynamics that we observe experimentally and describe theoretically using noisy simulations, matrix product states, and mappings to statistical mechanics models. Estimates of the universal critical properties are obtained to high accuracy on the quantum computer thanks to the large number of qubits utilized in the calculation. By successfully applying up to nearly 5000 entangling gates and 5000 non-unitary mid-circuit operations on systems up to 100 qubits, this experiment serves as a signpost on the route towards fault tolerance.
}
\end{abstract}
\maketitle


Chaotic systems~\cite{strogatz2001nonlinear} are ubiquitous in nature, manifesting in contexts as varied as turbulent weather~\cite{coiffier2011fundamentals} and irregular bus arrivals~\cite{krbalek2000statistical}. Yet within this apparent randomness, order often emerges: jet streams organize from turbulence~\cite{vallis2017atmospheric}, neural activity synchronizes into brain rhythms~\cite{buzsaki2006rhythms}, and predator–prey dynamics produce recurring population cycles~\cite{murray2002mathematical}. Beyond such naturally emergent orders, chaotic dynamics can 
be brought into order through control. The physical principles underlying chaos-control strategies involve continuous monitoring with feedback~\cite{PhysRevLett.64.1196}, time-delayed self-feedback~\cite{pyragas1992continuous}, and probabilistic control  protocols~\cite{antoniou1997probabilistic}. These ideas underpin important real world applications, from corrective pacing that suppresses cardiac arrhythmias~\cite{garfinkel1992controlling}, to feedback-stabilized laser dynamics~\cite{sivaprakasam2001experimental}, to small thrusts that guide spacecraft through unstable gravitational regions~\cite{koon2000heteroclinic}. Thus, the interplay between chaos and order is both a natural phenomenon and a guiding principle for controlling complex dynamical systems with broad applications.


The advent of quantum computing platforms---including superconducting qubits~\cite{kjaergaard2020superconducting}, trapped ions~\cite{bernardini2023quantum}, and neutral atoms~\cite{henriet2020quantum}---raises the fundamental and practical question of whether classical chaos-control methods can be adapted to control quantum dynamics. 
Controlling quantum chaotic dynamics~\cite{srednicki1994chaos, deutsch1991quantuma, rigol2008thermalization,d2016quantum,ippoliti2022solvable,zakrzewski2023quantum,choi2023preparing,PhysRevLett.130.020201,iadecola2023measurement} presents a host of challenges, as chaotic systems amplify small perturbations exponentially, making them highly responsive yet extremely fragile to noise, decoherence, and experimental imperfections. Chaotic energy spectra further complicate the use of standard resonance-based or adiabatic control strategies. 
The quantum extension of the aforementioned classical-chaos control strategies overcomes these difficulties by exploiting intrinsic features of chaotic phase space---most notably unstable periodic orbits---which serve as natural hot spots for control. This approach leverages ambient instabilities to enhance rather than undermine the stabilization of targeted states with control, thereby transforming chaos from an obstacle into a resource~\cite{PhysRevLett.64.1196} for robust quantum control.



In this work, we report the experimental realization of such quantum control on up to 100 qubits using an IBM superconducting quantum processor.
This is made possible by the rapid development of programmable quantum computers (QCs) over the last decade, which has enabled the synthesis of diverse quantum evolutions, including putatively chaotic ones, from random unitary gates~\cite{kandala2019error,kim2023evidence,robledo2025chemistry,yoshioka2025krylov,mi2021information,arute2019quantum}. 
Multiple QC platforms are beginning to go beyond unitary evolution by performing mid-circuit measurements (that only collapse the wavefunction locally)~\cite{pino2021demonstration,govia2023randomized,PhysRevLett.129.203602} while the unitary dynamics is underway to learn something about the system and potentially correct unwanted errors; this is a crucial step towards the realization of quantum error correction protocols.
The combination of unitary dynamics and projective measurements has led to the discovery of a measurement-induced phase transition (MIPT) in the entanglement structure of the wavefunction~\cite{skinner2019measurementinduced,li2018quantum,li2019measurementdriven, chan2019unitaryprojective,fisher2023random} and novel state preparation schemes to create and manipulate useful quantum states \cite{roy2020measurementinduced,puente2024quantum, barbarino2020preparing,smith2024constantdepth, pan2024topologicala, bhuiyan2025free}. 
Recently, this progress in performing midcircuit measurements has translated into using the outcome of mid-circuit measurements to perform a subextensive number of error correction rounds through real-time feedback on the quantum state~\cite{google2023suppressing,google2025quantum}. 
Developing quantum hardware that can quickly and accurately perform mid-circuit measurements and feedback is the crucial next step needed to move to the fault-tolerant era of quantum computing.

\begin{figure*}[htbp]
    \centering
     \includegraphics[width=\linewidth]{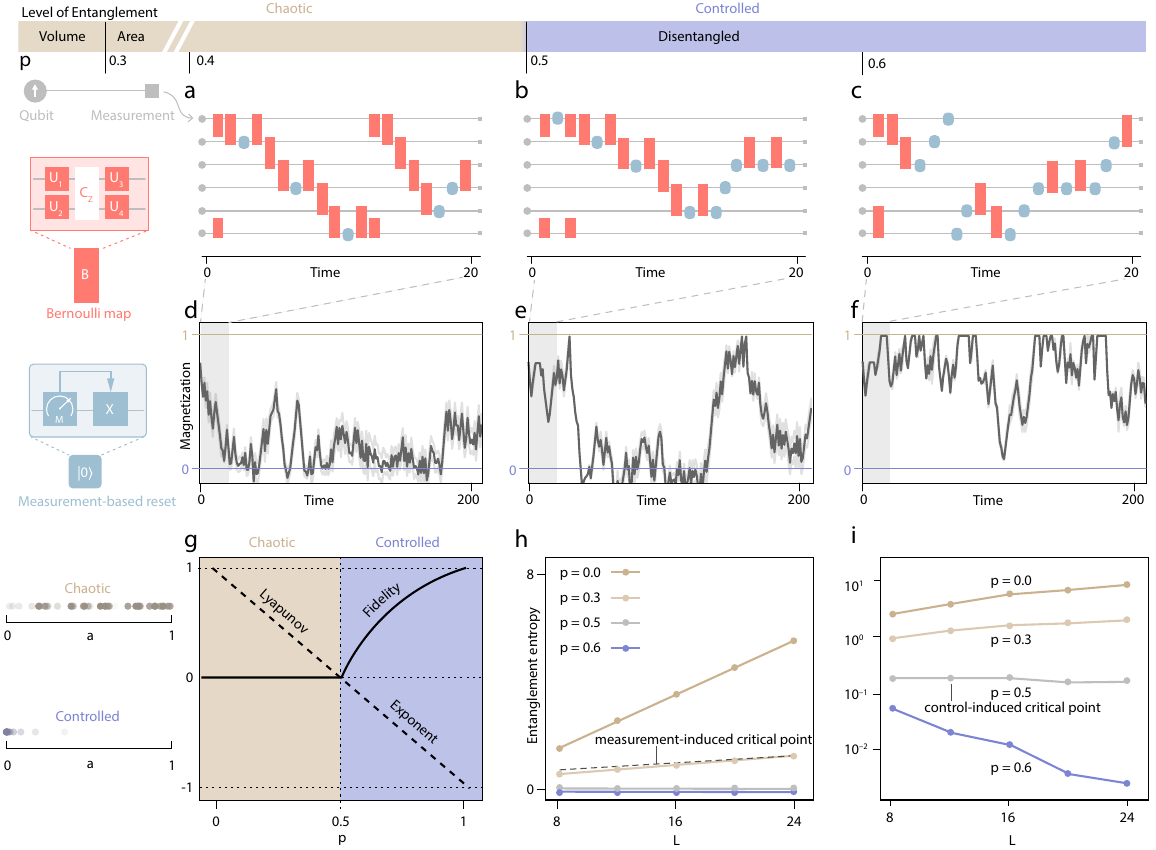}
    \caption{
    \textbf{Schematic for measurement and control induced phase transitions.} 
    Quantum circuit diagrams for the three regimes of the model we focus on are shown in terms of the properties of observables and quantum entanglement. The gates that sample the quantum Bernoulli map are shown as red rectangles (breakdown to the left of {\bf a}, see Methods), whereas the measurements and resets are blue squares (see left of {\bf d}). {\bf a}, {\bf b}, and {\bf c} correspond to typical circuits in the chaotic phase ($p=0.4$), the critical point ($p=0.5$), and the controlled phase ($p=0.6$), respectively. 
    {\bf d}--{\bf f} show time-series of the magnetization density $\langle M_z \rangle=\frac{1}{L}\sum_i \langle Z_i \rangle$ from each circuit diagram at the respective phases, where the shaded region is the standard error of mean estimated from 500 shots. 
    Left of {\bf g} depicts the action of the Bernoulli map ($a\rightarrow 2a$ mod 1) and control $(a\rightarrow a/2)$ on the number line (i.e., in the classical limit and $L\rightarrow \infty$). Dots depict one-time trace, and the darker dots are later time steps.
    {\bf g} shows the Lyapunov exponent $\lambda/\log 2 = (1-2p) $ for the Bernoulli map and the fidelity onto the fixed point $|\langle 0|\psi\rangle|^2$ in the chaotic and controlled phases.
    The Lyapunov exponent changes from positive to negative at the critical point $p=0.5$, as shown by the dashed line.
    {\bf h} and {\bf i} show the von Neumann entanglement entropy of the half-chain on linear and log scales, respectively, as the late-time entanglement transitions from a volume to an area law (within the chaotic phase) and then to an unentangled control phase at large enough $p$.
    }
    \label{fig:fig1}
\end{figure*}

To implement measurement- and feedback-assisted quantum control, we utilize a class of adaptive quantum circuits with a well-defined classical limit. This class of circuits stochastically interleaves dynamics associated with a quantum analog of the chaotic Bernoulli map~\cite{renyi1957representations} with a control operation that drives the quantum dynamics toward an unstable fixed point of the classical map~\cite{antoniou1996probabilistic,antoniou1997probabilistic,antoniou1998absolute}.

This family of quantum circuits includes random unitary gates interspersed with quantum measurements and resets over a time scale that scales with the number of qubits in the system.
Under local feedback operations, the model has been theoretically shown to possess an MIPT between extensive volume-law entanglement and subextensive area-law entanglement, and a further control-induced phase transition (CIPT) between this area-law phase and an essentially classical one where the dynamics ``sticks'' to the fixed point~\cite{pan2024local}, depicted in Fig.~\ref{fig:fig1}.
This model is useful as it retains access to the classical control protocols:
It allows for quantum fluctuations to be engineered into the chaotic phase by suitably choosing the gate set to induce a range of different superpositions of states---for example, stabilizer (or Clifford)~\cite{lemaire2024separate}  and fully generic quantum (i.e.,~Haar)~\cite{iadecola2023measurement,pan2024local} gates have been considered.
Recent theoretical efforts~\cite{pan2025controldriven} have demonstrated how to probe quantum fluctuations revealed by measurements and feedback, namely the fluctuations between different collapsed wavefunctions due to the probabilistic nature of quantum measurements. 
In the presence of measurements, even a circuit with a fixed structure admits a host of measurement outcomes, each one an independent realization of the collapse process.
The fluctuations of these measurement outcomes can be driven critical and in principle observed without the need for post-selection onto a fixed measurement history. 
However, the realization of this system in a quantum processor has remained well out of reach due to the extensive number of mid-circuit measurements and resets required.

We have overcome these challenges to implement this dynamics experimentally on an IBM superconducting quantum processor for system sizes up to 100 qubits and a number of quantum operations scaling quadratically with system size. 
The experimental data reveal critical properties at the CIPT with precise universal finite-size scaling behavior in both the dynamics and the long-time steady state of the system. 
Going beyond conventional averaging techniques, we further measure the quantum fluctuations that arise from the interplay of chaos, measurement, and feedback for the first time, showing that they become critical with universal critical exponents consistent with our theoretical calculations. 
Furthermore, we accurately compare the experimental distributions of quantum observables to noisy theoretical simulations with and without quantum fluctuations, certifying that the quantum computer is sampling a random quantum process rather than purely incoherent stochastic noise.
This work represents a significant advance in the accurate unison of measurement and feedback operations to design adaptive quantum circuits.


\vspace{3mm}
\noindent\textbf{\large{}Model, observables, and phase diagram}
\vspace{1mm}

We consider a stochastic quantum circuit on a system of $L$ qubits with open boundary conditions, defined such that the (unnormalized) wavefunction after $t$ time steps is given by
\begin{align}
\label{eq:model}
    \ket{\psi_t}_{\mathbf m_{\mathcal C}}=O_{i_t}\dots O_{i_2}O_{i_1}\ket{\psi_0}.
\end{align}
At the $k$th time step, a local quantum operation $O_{i_k}=B_{i_k}$ or $C_{i_k}$ is applied, with probabilities $1-p$ and $p$, respectively.
The unitary operator $B_{i_k}$ is 
a two-body ``scrambling" operation acting on qubits $i_{k}$ and $i_k+1$, itself chosen at random from a particular distribution
(see \cref{fig:fig1} and Methods).
$C_{i_k}$ is a one-body ``control" operation that resets the state of qubit $i_k$ to $\ket{0}$ -- the reset is implemented by first measuring the qubit (with outcome $m_{k}=\pm 1$), and then applying an $X$ gate if the post-measurement state is $\ket{1}$.
We denote the measurement record for circuit realization $\mathcal C$ by $\mathbf m_{\mathcal C}=\{m_k\}^t_{k=1}$, with the convention that $m_k=0$ if no measurement is performed at step $k$.
The scrambler $B_{i_k}$ both pushes the system toward volume-law entanglement (a proxy for quantum chaos) and randomizes the sequence of bits that would be revealed by quantum measurement (a proxy for classical chaos).
The control $C_{i_k}$ opposes the scrambler by attempting to push the system toward the polarized state $\ket{0\dots 0}$ where all qubits are in their ground state.
The location $i_{k}$ on which these operations are applied is chosen according to a random walk with bias $p$: if $B_{i_k}$ is applied to site $i_k$ at the $k$th step, then $i_{k+1}=i_k+1$, while $i_{k+1}=i_k-1$ if $C_{i_k}$ is applied.
Example circuits for this protocol are shown in \cref{fig:fig1}{\bf a-c} for different values of $p$.

We call this stochastic quantum circuit the adaptive Bernoulli circuit model~\cite{iadecola2023measurement,lemaire2024separate,allocca2024statistical,pan2024local,pan2025controldriven} for its close connection to the classically chaotic Bernoulli map under stochastic control~\cite{antoniou1996probabilistic,antoniou1997probabilistic,antoniou1998absolute}.
In particular, when the initial state $\ket{\psi_0}$ is a computational basis state represented by a (classical) bit string and the random quantum gate $B_i$ is replaced by a random Markovian update of the bits at sites $i$ and $i+1$, the resulting dynamics is equivalent in a precise sense~\cite{iadecola2023measurement,lemaire2024separate} to that of the Bernoulli map $B(a) = 2a\text{ mod }1$ in stochastic competition with the control map $C(a)=a/2$.
In the classical model, $a\in[0,1)$ can be any (rational or irrational) number; the analogous circuit model on bit-strings captures the dynamics when $a$ is truncated at $L$ bits and converges to the original map as $L\to\infty$ (depicted in the left of \cref{fig:fig1}{\bf g}).

\begin{figure*}
    \centering
    \includegraphics[width=6.8in]{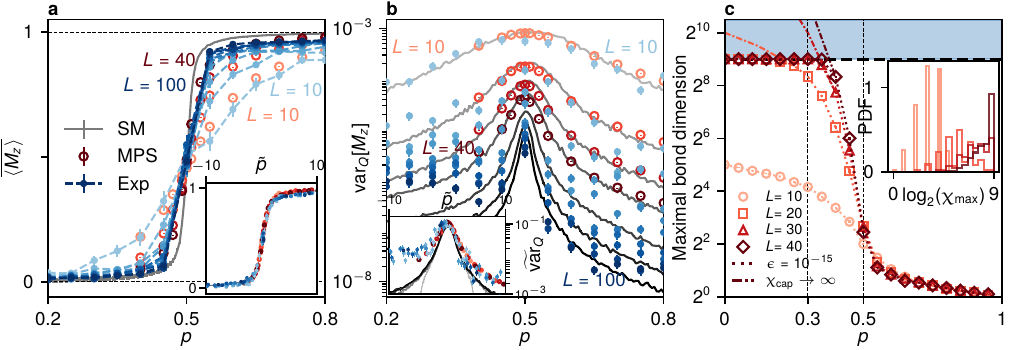}
    \caption{
    \textbf{Quantum-to-classical control transition in the long-time steady state.} Results are shown for experimental data from \texttt{ibm\_fez} (blue dots; $L=10$--$100$), matrix product state calculations (red circles; $L=10$--$40$), and a  statistical mechanics model (grey lines; up to $L=100$).
    \textbf{a} The magnetization density averaged over initial states, measurement outcomes, and circuits (denoted $\overline{\langle M_z \rangle}$) as a function of the control probability $p$ for various system sizes $L$. 
    (Inset) Data collapse of the magnetization as a function of $\tilde{p}=L^{1/\nu} (p-p_c)$ with: 
    $p_c=0.495(1)$, $\nu=1.05(2)$ for experimental data; 
    $p_c=0.4947(5)$, $\nu=1.05(2)$ for MPS calculations.
    $p_c=0.4970(5)$, $\nu=1.0000(2)$ for theoretical first moment stat mech model.
    \textbf{b} The variance over quantum fluctuations of the magnetization~\cite{pan2025controldriven} (see Eq.~\eqref{eq:quantum_fluctuation_Mz} in Methods) as a function of the control probability $p$ for various system sizes $L$. 
    (Inset) Finite-size scaling of the (scaled) variance, 
    $\widetilde{\text{var}}_Q = \Var{Q}{M_z} L^{2\beta/\nu}$ as a function of $\tilde{p}=L^{1/\nu} (p-p_c)$ with $p_c=0.509(2)$, $\nu=0.95(1)$, and $\beta=1.05(2)$ for experimental data;
    $p_c=0.500(6)$, $\nu=0.95(1)$,  and $\beta=1.01(1)$ for theoretical MPS calculations;
    and $p_c=0.5010(1)$, $\nu=1.00(2)$,  and $\beta=1.01(1)$ for theoretical first moment stat mech model.
    \textbf{c}  A proxy for the computational cost of the steady state represented through the {average maximal} bond dimension of the matrix product states used to describe the steady state wavefunction to an accuracy $\epsilon$ (set by the truncation error),
    as a function of $p$, at various $L$ and with a maximal bond dimension of $512$. 
    The bond dimension reflects the nature of the entanglement of each phase in \cref{fig:fig1} and the area law nature of the control transition.
    The dashed lines in the shading region are guides to the eye that smoothly interpolate between finite $p$ and its value at $p=0$.
    The two vertical dashed lines separate the volume-law phase, area-law phase, and disentangled phase.
    Inset: The distribution of the maximal bond dimension $\chi$ near the control transition $p=0.4$ inside the area law for $L=10$ to 40 (light to dark red).
    }
    \label{fig:fig2}
\end{figure*}

We realize the model~\eqref{eq:model} experimentally using the 156-qubit IBM superconducting processor \texttt{ibm\_fez} {(see Supplemental Material for devices' details)}.
To reach the steady state, we evolve for $t=L^2/2$ time steps, corresponding to a circuit volume $\sim L^2$, for $L$ up to 100 qubits.
This experiment takes advantage of high-fidelity measurement-based mid-circuit reset operations.
These experimental results are compared to simulations using matrix product states (MPSs), as well as noisy simulations assuming different types of error channels.
A further basis for comparison is provided by a series of statistical mechanics models (see Methods) that allow us to predict via simulations the average values of certain observables and their \textit{quantum} fluctuations.
Comparing the experimental data and theoretical calculations allows us to benchmark the accuracy of results from the quantum computer, and crucially enables an observed distinction between experimental reality and theoretical simulations.

The dynamics of the model \eqref{eq:model} can be probed using several observables.
The most straightforward of these is the magnetization density $M_z=\frac{1}{L}\sum^L_{i=1}\hat Z_i$, where $\hat Z_i$ is the Pauli-$Z$ operator on site $i$.
At large $p$, the control operation is applied frequently and the system always reaches the target state $\ket{0\dots0}$ at late times, corresponding to $\expval{M_z}\to 1$ as $t,L\to\infty$.
At small $p$, the scrambler dominates, leading to $\expval{M_z}\to 0$ in the same limit, see \cref{fig:fig1}{\bf d-f}.
A transition between these controlled and chaotic phases occurs at $p=0.5$, which has been verified theoretically as a robust feature of the quantum and classical versions of the model. At the transition from chaos to control, the Lyapunov exponent $\lambda=(1-2p)\log2$ changes sign, going from exponentially diverging trajectories to absorbing ones, as shown in \cref{fig:fig1} {\bf g}.
This transition is accompanied by a non-zero fidelity of the quantum wavefunction with the target state $\ket{0\dots0}$ (\cref{fig:fig1}{\bf g}).

The above features appear in both the classical
Bernoulli map with control and its quantum analog.
However, in the quantum setting, the qubit resets in the control map drive a phase transition in the entanglement of the late-time wavefunction.
Like the MIPT, the system exhibits extensive volume-law entanglement at small $p$ and zero entanglement at large $p$.
However, a transition from volume-law to subextensive area-law entanglement scaling occurs for $p=p_{\rm ent}\approx 0.3$~\cite{pan2024local}, well before the control transition (see \cref{fig:fig1} top). 
At this entanglement phase transition, the von Neumann entanglement entropy scales logarithmically with $L$ as in the standard MIPT [see \cref{fig:fig1}{\bf h}].
In contrast, at the control transition $p=0.5$, the entanglement entropy is constant as a function of $L$ and tends to zero as $L\to\infty$ for $p>0.5$ [see \cref{fig:fig1}{\bf i}].
Thus, the model \eqref{eq:model} has three distinct phases: a volume-law chaotic phase for $0\leq p < p_{\rm ent}$, a chaotic phase with area-law entanglement for $p_{\rm ent}< p < 0.5$, and an unentangled controlled phase for $0.5< p\leq 1$, shown in \cref{fig:fig1}{\bf i}. 
The area-law chaotic phase is particularly interesting: while the system is only modestly entangled, the wavefunction retains exponentially many superposition components~\cite{pan2025controldriven}.
Thus, the control transition at $p=0.5$ is akin to a quantum-to-classical transition. In the following, we directly observe this dynamical quantum-to-classical phase transition on IBM's superconducting processor.

\vspace{3mm}
\noindent\textbf{\large{}Experimental results}
\vspace{1mm}

In \cref{fig:fig2}, we examine properties of the long-time steady state of the dynamics (i.e., after $L^2/2$ operations).
\Cref{fig:fig2}{\bf a} shows the experimental measurement of the average magnetization $ \overline {\langle M_z \rangle}$ in the steady state as a function of the control rate $p$ for several different system sizes $L=10$-$100$ (solid blue symbols), displaying the quintessential behavior of each phase. The quantum average over the wavefunction in Eq.~\eqref{eq:model} is denoted $\langle \dots \rangle$, and the overline denotes an average over initial states, measurement outcomes, and circuit realizations.
For each value of the control rate $p$, fifty circuit realizations $\mathcal C$ are evaluated on the device with a fixed initial state (we have verified that the initial state does not affect the steady state).
Observables for each circuit realization are calculated by averaging over $10^4$ shots, each with a different (unknown) measurement history $\mathbf m_{\mathcal C}$.
No error mitigation is used to obtain the experimental results in \cref{fig:fig2}.
Noiseless theoretical calculations using matrix product states (open red symbols) for sizes $L=10$-$40$ with $p\ge 0.4$ and using a statistical mechanics model (grey lines) agree well with the experimental data (see Methods for more details). 
The experimental and theoretical data all collapse onto a universal curve following $ \overline {\langle M_z \rangle}\sim f_M[(p-p_c)L^{1/\nu}]$ (\cref{fig:fig2}{\bf a} inset). 
From the data collapse we find $p_c=0.495(1)$ and $\nu=1.05(2)$ for the experimental data, as well as $p_c=0.4950(5)$ and $\nu=1.05(2)$ for the MPS calculations, and $p_c=0.4970(5)$, $\nu=1.0000(2)$ for the statistical mechanics model signaling excellent agreement between the quantum experiments and the theoretical predictions.

The realization-averaged magnetization has similar behavior to the average magnetization in the classical Bernoulli circuit model~\cite{iadecola2023measurement}. 
To experimentally probe the quantum nature of the control transition we therefore turn to considering the quantum fluctuations by measuring the variance of the magnetization over measurement outcomes for a fixed circuit $\mathcal C$, which we denote by $(\Delta M_z)^2\vert_{\C}$
(see Methods for the precise definition and Sec.~\ref{sec:fss} in the Supplemental Material for the finite size scaling to obtain the critical exponents). 
In this context, the variance of any observable can be decomposed into a sum of classical and quantum contributions~\cite{pan2025controldriven}, and we focus on a global observable $M_z$ in the following to reduce the effects of noise (see Methods for precise details).
The quantum contribution to the variance originates from the randomness of the quantum measurement outcomes (i.e., the fluctuations between different wavefunction collapses) and is precisely zero in the classical limit where measurement outcomes are deterministic~\cite{pan2025controldriven}.
These quantum fluctuations can be measured from the distribution $\mathcal{P}$ of $(\Delta M_z)^2\vert_{\C}$ over different circuit realizations $\mathcal{C}$.
In \cref{fig:fig2}{\bf b}, we plot the variance over circuits $\mathcal C$ of the quantum fluctuations (i.e. the second central moment of $\mathcal{P}$), denoted $\Var{Q}{M_z}=\Var{\C}{(\Delta M_z)^2\vert_{\C}}$,
as a function of the control rate $p$.
This is done for the experimental data (solid blue symbols), noiseless matrix product state simulations (open red symbols) and the statistical mechanics model (grey lines). 
Near $p_c$ we find excellent data collapse for both experimental data and the theoretical calculations as shown in \cref{fig:fig2}{\bf b} (inset), displaying the universal scaling $\Var{Q}{M_z}\sim L^{-2\beta/\nu}f_Q[(p-p_c)L^{1/\nu}]$ with $p_c=0.509(2)$, $\nu=0.95(1)$, and  $\beta=1.05(2) $ from the experimental data,   $p_c=0.500(6)$, $\nu=0.95(1)$, and  $\beta=1.01(1)$ for theoretical MPS calculations, and $p_c=0.5010(1)$, $\nu=1.00(2)$,  and $\beta=1.01(1)$ in the statistical mechanics model, once again demonstrating very good agreement with the experiment.

To gauge the complexity of the simulations achieved experimentally on the quantum computer in comparison with established computational techniques, in \cref{fig:fig2}{\bf c} we plot theoretical data on the maximum bond dimension $\chi_{\rm max}$ needed to achieve a fixed error threshold $\epsilon=10^{-15}$ for $L=10$--$40$ as a function of $p$.
The bond dimension $\chi$ is a quantitative metric for matrix product state simulations that controls the amount of entanglement accessible to the simulation (see Methods).
Matrix product states with larger bond dimension can accommodate more entanglement, but require more computational resources to store and manipulate.
For $p<0.3$, below the MIPT into the area-law phase, the bond dimension reaches the numerical cutoff $\chi_{\rm cap}=512$ for all but the smallest system size.
The dot-dashed lines indicate extrapolations beyond $\chi_{\rm cap}$, showing the increasing growth rate of $\chi_{\rm max}$ with decreasing $p$ as $L$ increases.
In fact, even in the area-law part of the chaotic phase ($0.3 < p < 0.5$), the bond dimension is not yet saturated in $L$ except very close to the transition, even though it is expected to eventually reach an $L$-independent value for sufficiently large $L$.
Further insight can be found in the inset, which plots the distribution of $\chi_{\rm max}$ over realizations of the dynamics within the area law phase at $p=0.4$---crucially, we see that a significant number of realizations hit the cutoff $\chi_{\rm cap}$ as $L$ increases.
This is because individual circuit realizations can still produce volume-law states that are challenging to simulate classically, even within a phase where typical quantum trajectories are area-law.
Thus, in the volume-law phase and even part of the area-law phase, classical simulations have difficulty producing accurate results for large $L$, while the quantum computer efficiently explores the full phase diagram as shown in \cref{fig:fig2}{\bf a},{\bf b}.

\begin{figure*}
    \centering
    \includegraphics[width=6.8in]{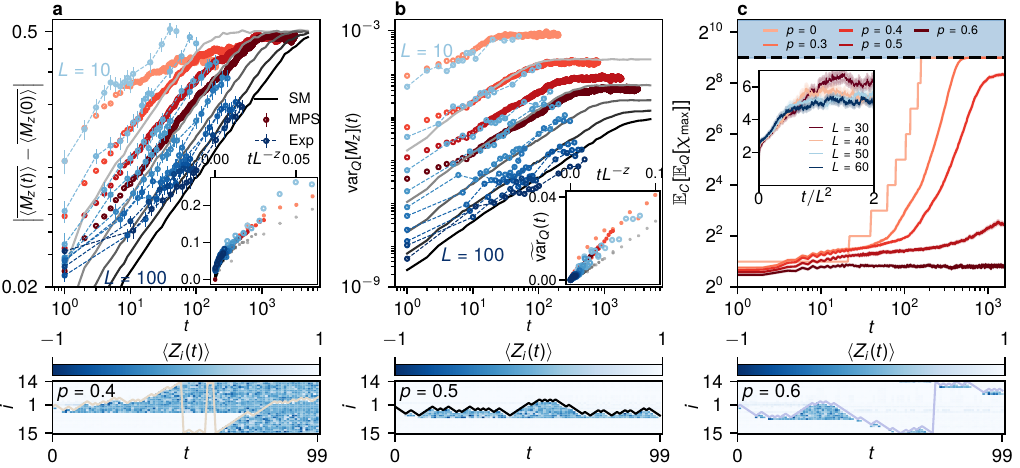}
    \caption{
    \textbf{Dynamical nature of the control transition.} 
    Results are shown for experimental data from \texttt{ibm\_fez} (blue dots; $L=10-100$), matrix product state calculations (red circles; $L=10-40$), and a statistical mechanics model (grey lines; up to $L=100$).
    {\bf a} At the control transition $p=0.5$ we plot the dynamics of the magnetization density averaged over initial states, measurement outcomes, and circuits as a function of time, $\overline{\langle M_z(t) \rangle}$ (subtracting off its initial value), showing its growth out to times $t=2L^2$ for MPS simulation, $t=L^2/2$ for stat mech model, and $t=5L$ for experimental data. 
    (Inset) Data collapse of the growth of $\overline{\langle M_z(t) \rangle}-\overline{\langle M_z(0) \rangle}$ versus $t/L^z$ yielding the consistent estimates of the dynamical exponent $z=2.00(1)$ from experimental data, $z=2.00(2)$ from matrix product state calculations, and $z=2.000(2)$ from first moment stat mech model.
    {\bf b} Dynamics of the quantum fluctuations of the magnetization variance~\cite{pan2025controldriven} (see Eq.~\eqref{eq:quantum_fluctuation_Mz} in Methods), as a function of times out to the same final times as in {\bf a}.
    (Inset) Data collapse following $\widetilde{\text{var}}_{Q}(t) = \Var{Q}{M_z}(t)L^{2\beta}$ vs. $t/L^z$ yields $z=2.08(1), \beta = 1.00(6)$ from the experimental data, $z=2.001(3)$ and $\beta = 1.05(1)$ from matrix product state calculations, {$z=2.00(3)$ and $\beta = 0.95(7)$ from the first-moment stat mech model. The data collapse range is the early times of the growth with $t\le L$.}
    \textbf{c} Growth of the {average} bond dimension of the matrix product states used in the simulations for $L=40$ at different control probabilities $p$. (Inset) At the control transition $p=0.5$, the collapse of the growth of the bond dimension with $t/L^2$.
    (Bottom panels) The dynamics of the experimental local averaged magnetization given a specific circuit at $L=30$ and $p=0.4, 0.5$, and 0.6, respectively, from left to right panels, where the first qubit is cycled to the middle of the system for better visibility. The pink/black/blue lines correspond to the position where the Bernoulli and the control map apply in a single realization of the circuit, showing a biased/unbiased/biased random walk behavior, respectively. 
        }
\label{fig:fig3}
\end{figure*}

In \cref{fig:fig3} we turn to the dynamical behavior of the circuit at the control transition $p=p_c=0.5$.
The bottom panels of \cref{fig:fig3} highlight the nature of the dynamics below, at, and above the control transition ($p=0.4, 0.5,$ and $0.6$, respectively) through the site-resolved dynamics of the local magnetization.
The local magnetization dynamics transitions from scrambling to nonscrambling dynamics as $p$ increases.
For $p<0.5$, the front of scrambled qubits moves leftward and eventually wraps around the system, while for $p>0.5$ the front moves rightward and preempts scrambling from consuming the system.
At $p=0.5$, the front executes an unbiased random walk, which leads to telltale diffusive scaling.

\Cref{fig:fig3}{\bf a} shows experimental data (solid blue symbols) for the realization-averaged magnetization as a function of time $t$, $\overline{\expval{M_z(t)}}$, at system sizes $L=10$--$100$.
In particular, we plot the change of the realization-averaged magnetization from its initial value at $t=0$, observing a clear power-law growth.
The data for different system sizes collapse according to $\overline {\langle M_z(t) \rangle}\sim f'_{M}(t/L^z)$, with a dynamical exponent $z=2.00(1)$ (see inset), consistent with the expected value $z=2$ associated with the diffusive random-walk behavior.
For comparison, we show results from MPS simulations (open red symbols) for $L=10$--$40$ and theoretical results from the first-moment stat mech model (grey lines; see Supplemental) that collapse with  $z=2.00(2)$ and $z=2.000(2)$, respectively, demonstrating excellent agreement with the experiment.

\Cref{fig:fig3}{\bf b} shows experimental results for the dynamics of the quantum fluctuations $\Var{Q}{M_z}(t)$ at the control transition.
These also exhibit a clear power-law growth, with scaling collapse to the universal form $\Var{Q}{M_z}(t)\sim L^{-2\beta/\nu}f'_Q[t/L^z]$ (inset), revealing a dynamical exponent $z=2.08(1)$ and scaling dimension $\beta=1.00(6)$ for the experiment.
Theoretical results from MPS simulations display collapse with very similar exponents $z=2.001(3)$ and $\beta=1.05(1)$, and those from the first-moment stat mech model also show similar exponents $z=2.00(3)$ and $\beta=0.95(7)$.
We emphasize that while the classical critical properties control this behavior, these fluctuations originate from quantum processes.

\Cref{fig:fig3}{\bf c} shows theoretical results for the growth of the maximal bond dimension as a function of time for $L=40$ and five different values of $p = $ 0 to 0.6 (from light red to dark red shown above the dashed line).
In the volume-law phase $p<0.3$, the bond dimension grows exponentially with $t$ until it hits the arbitrary cutoff $\chi=512$.
Theoretically, it would ultimately saturate to the maximum value $2^{L/2}$, signaling the intractability of simulating the dynamics in this regime with MPS techniques.
In the area-law chaotic regime $0.3<p<0.5$, it increases over time but eventually saturates to an area-law value.
Finally, deep in the controlled phase $p>0.5$, it hardly grows at all with time. At the transition, the bond dimension grows diffusively following the random walk universality and grows like $t/L^2$ (\cref{fig:fig3}{\bf c} inset).


\vspace{3mm}
\noindent\textbf{\large{}Discussion}
\vspace{1mm}

\begin{figure*}
    \centering
    \includegraphics[width=\linewidth]{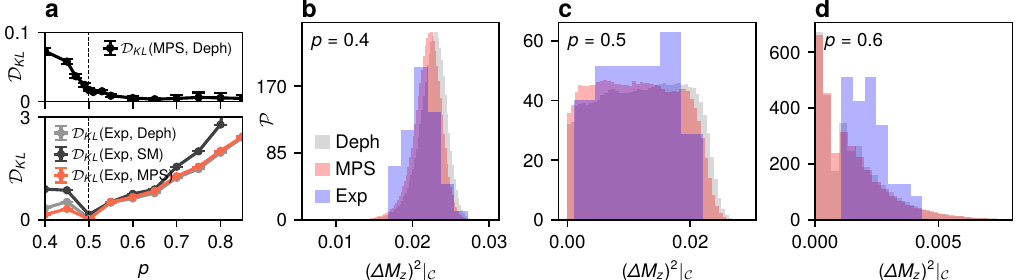}
    \caption{
        \textbf{Distribution of the quantum fluctuation of magnetization from the experiment, stat mech model, and dephasing model, and their Kullback-Leibler divergence.}
        \textbf{a} Top panel: Kullback-Leibler (KL) divergence between distribution of quantum fluctuation of the magnetization $(\Delta M_z)^2\vert_{\C}$ from the MPS and dephasing model (Deph); Bottom panel: KL divergence between the distribution of quantum fluctuation of the magnetization $(\Delta M_z)^2\vert_{\C}$ from the experiment (Exp) and the MPS (red line), between that from experiment and the dephasing model (light gray line), and between that from the experiment and the first-moment statistical mechanics (SM) model (dark gray line).
        The error bar is the bootstrapping in the kernel density estimate with 100 times and $10^4$ sample sizes.
        \text{b-d} Probability density function $\mathcal{P}$ of the quantum fluctuations of the magnetization $(\Delta M_z)^2\vert_{\C}$ for the experiment (blue), MPS (red), and the dephasing model (gray) at $p=0.4$, $p=0.5$, and $p=0.6$.
    }
    \label{fig:fig4}
\end{figure*}

The quantum fluctuations plotted in \cref{fig:fig2}{\bf b} and \cref{fig:fig3}{\bf b} would be \textit{identically zero} in the classical limit of this problem. 
However, in the interest of certifying the ``quantumness'' of the experimental results from the quantum processor, it is worth identifying, as points of comparison, noisy classical processes that could ``spoof'' these fluctuations.
To do so, we can replace the quantum gates with a classical Markov process for dephasing errors. 
This Markov process can be thought of as a very noisy quantum computer where a bath measures the qubits in the computational basis immediately after a gate is performed, thereby killing entanglement and coherent quantum fluctuations and replacing them with incoherent statistical fluctuations. 
This is distinct from the statistical mechanics model, which is a precise simulation of certain quantum observables, e.g., the variance of the magnetization. 
This dephasing model simulates a maximally noisy quantum computer in the dephasing noise channel, while the previous statistical mechanics model in Figs.~\ref{fig:fig2} and~\ref{fig:fig3} can be considered as a depolarizing channel, which is most destructive to bit-string information. 

To compare this with the experiment, in Fig.~\ref{fig:fig4} we analyze the distribution of the magnetization in each case and compute the Kullback-Leibler (KL) divergence (see Methods and Sec.~\ref{sec:KL} in the Supplemental Material) of these distributions, which is non-zero if the distributions are distinct.
This allows us to compare two separate theoretical models with the experimental data:
(i) the MPS simulations of the quantum dynamics, and
(ii) the dephasing noisy Markov process.

The quantity that captures this well is $(\Delta M_z)^2\vert_{\C}$,
i.e., the fluctuations in the magnetization for a given circuit realization.
To analyze the distribution of these fluctuations, we use {50} different circuit realizations and {1000} runs of the experiment per realization. Each point represents a variance over shots within the same circuits to plot the histogram. 
With this, we can compare the experimental distribution of $(\Delta M_z)^2\vert_{\C}$ to the distributions generated by the dephasing model and the fully quantum MPS simulations. 
In \cref{fig:fig4}{\bf a} we see four comparisons: The black curve (top panel) represents the KL divergence between the MPS simulations and the dephasing model, while the red and grey curves (bottom panel) represent the KL divergence between the experimental data and the MPS and dephasing models, respectively.
For $p\leq p_c$, the black curve is nonzero, indicating that the distributions of fluctuations generated by the MPS simulations and the dephasing model are distinct.
This means that, in principle, the  dephasing model cannot spoof the distribution of fluctuations in the fully quantum model.
In contrast, as $p$ increases, the black curve monotonically decreases to zero, consistent with the intuition that the system behaves essentially classical in the controlled phase. 
(These trends are amplified with increasing system size, see Sec.~\ref{sec:KL} in the Supplemental Material).

The red, light grey, and dark grey curves can be used to discern whether the experimental data is more consistent with the MPS simulations, the dephasing model, or the first-moment statistical mechanics model, respectively, with smaller KL divergence indicating closer agreement. Here, the first-moment statistical mechanics model (dark gray curve) although provides a good description of the critical point in Figs.~\ref{fig:fig2} and~\ref{fig:fig3}, it fails to capture the full distribution of fluctuations inside chaotic or controlled phases, and is also worse than the dephasing model in describing the experimental data as a ``classical model''.
A few features are of note: For $p>0.5$, noise begins to dominate the experimentally measured fluctuations, and we see none of the theoretical model captures the experiment.
However, for $p<0.5$ we notice that the experiment is better described by the MPS simulations than by the dephasing model (across all problem sizes, see Sec.~\ref{sec:KL} of Supplemental Material).
The crucial point here is that the distribution of $(\Delta M_z)^2\vert_{\C}$ is distinct for classical and quantum versions of this simulation, and the experiment is in better agreement with the quantum model than any stochastic and incoherent model that reasonably captures the system.
To see this qualitatively, \cref{fig:fig4}{\bf b-d} show the full distributions for the experiment (blue), the MPS simulations (red), and the dephasing model (grey) for $p=0.4, 0.5,$ and $0.6$, representing the chaotic, critical, and controlled regimes, respectively.
The theoretical distributions for the dephasing model and quantum model (grey and red) show slight differences in the chaotic phase and at criticality.
While it is a subtle distinction, the experimental data (blue) is slightly more in line with the quantum distributions than the dephasing model's distributions, as confirmed quantitatively by the KL divergence.
In a word, a noisy classical process (the dephasing model) does not fit the data as well as a full quantum simulation.

We should note, in \cref{fig:fig4}{\bf d}, that the strong peak at zero variance is not present in the experimental data. 
This is chiefly due to device noise. 
In the ideal, noiseless case, the peak at zero should correspond to the entire system being in the all-zero state. 
When there is an unmitigated readout error, the system is never exactly in the all-zero state, which is what leads to the deviation at zero variance that results in the finite KL divergence for $p>p_c$ in \cref{fig:fig4}{\bf d}. 
The noisy data still captures some of the tail of the distribution, just not the peak at exactly zero variance.
This explains why, on the controlled side, neither the dephasing model nor the MPS captures the experimental data. 
\vspace{3mm}
\noindent\textbf{\large{}Conclusion and outlook}
\vspace{1mm}

In this work, we have experimentally demonstrated the coherent control of up to 100 superconducting qubits to probe a novel far-from-equilibrium dynamical quantum phase transition from chaotic to controlled---and from quantum to effectively classical---behavior.
The experiment accesses the highly entangled volume-law phase of the model in addition to the more classically tractable area-law phase, and produces quantitatively accurate critical properties of the control transition.
Moreover, it observes for the first time the fluctuations of the order parameter between quantum trajectories at the phase transition, as well as their critical properties, which are not present in previously studied nonequilibrium phase transitions~\cite{Chertkov2023}.
Finally, we also certify that the experimental results are better described by the MPS simulations of the quantum model (at system sizes accessible to the numerics) than by an effective classical model (stochastic or deterministic) that washes out the quantum coherence.

This work represents a substantial step forward in the study of nonequilibrium quantum many-body physics with quantum computers.
Previous experimental realizations of MIPTs on trapped-ion~\cite{noel2022measurementinduced} and superconducting~\cite{Koh2023} quantum processors were limited to tens of qubits and mid-circuit measurements. 
A teleportation transition was observed in Ref.~\cite{hoke2023measurementinduced} on up to 70 qubits using a space-time duality mapping to avoid mid-circuit measurements altogether and maintain a circuit depth of order one.
Recent work examining quantum and classical nonequilibrium phase transitions on a trapped-ion quantum processor used 20 physical qubits to simulate systems of up to 73 qubits over 18 rounds of evolution, including qubit resets~\cite{Chertkov2023}.
In contrast, here we realize quantitatively accurate quantum simulations on up to 100 physical qubits with evolutions containing up to $10^4$ unitary gates and mid-circuit measurement-based resets (see Sec.~\ref{sec:experimental_details} in the Supplemental Material for more details).
Such coordination between unitary gates, measurement, and feedback is a crucial ingredient to reach fault tolerance, and this work demonstrates that results of scientific interest can be generated with present-day quantum computers in advance of this ultimate goal~\cite{kim2023evidence}.



\vspace{3mm}
\noindent\textbf{\large{}Methods}
\vspace{1mm}


\textbf{Device Specifications} 
Our experiments utilized \verb|ibm_fez|, a Heron r2 processor comprising 156 fixed-frequency transmon qubits. 
For each experimental realization, 100 linearly connected qubits out of the 156 qubits were used. These qubits were chosen by considering the device parameters, in particular the relaxation time $T_1$, the decoherence time $T_2$, measurement errors, single-qubit gate errors, and the two-qubit gate errors, to maximize the expected fidelity of the quantum circuits. The device's specifications are detailed in Sec.~\ref{sec:device_specifications} in Supplemental Material.

\textbf{Choice of approximated Haar random unitaries in the experiment}\label{sec:approx_Haar}
In the experiment, to reduce the noise rate as much as possible, we approximate the two-qubit Haar random unitary acting on qubit $ i$ and $i+1$, as shown in the red rectangle in \cref{fig:fig1}{\bf a-c}, with single-qubit rotations coupled by a two-qubit CZ gate (if $i$ and $i+1$ are nearest neighbours), namely,
\begin{equation}\label{eq:B}
    B_{i,i+1} = [\hat{U}_1\otimes \hat{U}_2 ] [\CZ_{i,i+1}]^{1-\delta_{i,L}} [\hat{U}_3\otimes \hat{U}_4 ],
\end{equation}
where $\hat{U}_1,\hat{U}_3$ acting on qubit $i$ and $\hat{U}_2,\hat{U}_4$ acting on qubit $i+1$, as shown in the red rectangle left of \cref{fig:fig1}{\bf a}, and $\delta_{i,j}$ is the Kronecker delta function.
Here, each single-qubit rotation is composed of single-qubit rotations around the $x$ and $z$ axes, i.e.,
\begin{equation}\label{eq:approx_Haar}
    \hat{U}(\theta,\vartheta,\Theta)= \hat{R}_x(\theta)\hat{R}_z(\vartheta)\hat{R}_x(\Theta),
\end{equation}
where $\theta,\vartheta,\Theta$ are independently drawn from a uniform distribution on $[0,2\pi)$, and $\hat{R}_\alpha(\theta)=\exp(-\ii\frac{\theta}{2}\hat{\sigma}_\alpha)$ is the rotation gate around the $\alpha$ axis with $\alpha=\left\{ x,z \right\}$ and $\hat{\sigma}_\alpha$ being the corresponding Pauli matrix.
Although the parametrization of $B_{i,i+1}$ in \cref{eq:B}, giving rise to a random ensemble $\mathcal{D}$, does not recover the uniform Haar random distribution, it captures the essential features of the chaotic dynamics, and also is very close to the Haar random unitary as the frame potential \cite{harrow2009random} is numerically found to be $\approx 2.14$, which is close to the value of $2$ for the Haar random unitary.

\textbf{Quantum shot-to-shot fluctuation of the magnetization}
In this section, we provide the definition of the quantum shot-to-shot fluctuation over an observation (e.g., magnetization $M_z$) as introduced in Ref.~\cite{pan2025controldriven}.
In the noise-free setting, the total fluctuation of any observable $\mathcal{O}$ is $\sigma^2[\mathcal{O}]=\E{\C}{\E{\mC}{\mathcal{O}_{\mC}^2}} - \E{\C}{\E{\mC}{\mathcal{O}_{\mC}}}^2$, where $\mathcal{O}_\mC$ is just one shot from any random circuit and mid-circuit measurement outcome.  This total fluctuation conceptually contains all the sources of randomness, including the randomness from the choice of the circuit $\C$ and the randomness from the measurement outcomes $\mC$ for a given circuit, and the final wave function collapse, where the first source of randomness is ``classically'' controllable while the latter two are ``quantum'' random intrinsically.

Therefore, we can decompose this total fluctuation into two parts, the circuit fluctuation and the quantum fluctuations, as per
\begin{equation}
    \begin{split}
        \sigma^2[\mathcal{O}] =& \underbrace{\mathbb{E}_{\mathcal{C}} \left[ \left(\mathbb{E}_{\bm{m}_{\mathcal{C}}}\left[O_{\bm{m}_{\mathcal{C}}} \right] \right)^2 \right] - \left( \mathbb{E}_{\mathcal{C}} \mathbb{E}_{\bm{m}_{\mathcal{C}}} \left[O_{\bm{m}_{\mathcal{C}}} \right] \right)^2}_{\text{circuit fluctuations}}\\
        +&\mathbb{E}_{\mathcal{C}}\underbrace{ \left[\mathbb{E}_{\bm{m}_{\mathcal{C}}} \left[O^2_{\bm{m}_{\mathcal{C}}} \right] - \left( \mathbb{E}_{\bm{m}_{\mathcal{C}}} \left[O_{\bm{m}_{\mathcal{C}}} \right] \right)^2 \right]}_{\text{quantum fluctuations}}.
    \end{split}
\end{equation}
In this work, we focus on the quantum shot-to-shot fluctuation $(\Delta M_z)^2\vert_{\C} = \E{\mC}{ \ev{M_z^2}{\psi_{\mC}} } - \left( \E{\mC}{\ev{M_z}{\psi_{\mC}}} \right)^2$, and refer the reader to Ref.~\cite{pan2025controldriven} for more details on the circuit fluctuation.

In principle, the fundamental definition for the order parameter showing the classical-to-quantum transition is the weight of zero fluctuation $P((\Delta M_z)^2\vert_{\C}=0)$.
However, due to the intrinsic noise in the quantum device, which always leads to a finite variance of the magnetization density, smearing the criticality of $P((\Delta M_z)^2\vert_{\C}=0)$, we use an alternative order parameter as a workaround, which goes beyond previous considerations~\cite{pan2025controldriven} to investigate the circuit dependence of the inherent quantum fluctuations of the system, denoted as 
\begin{equation}\label{eq:quantum_fluctuation_Mz}
    \Var{Q}{M_z} = \mathrm{var}_{\C}\left[ { (\Delta M_z)^2\vert_{\C} } \right],
\end{equation}
Given the dimension of the quantum fluctuation of the magnetization carries a finite scaling dimension $\beta$, the variance of it should be scaled as $\Var{Q}{M_z} \sim L^{-2\beta/\nu} f_Q[(p-p_c)L^{1/\nu}]$, as shown in \cref{fig:fig2}{\bf b}.

\textbf{Finite Size Scaling Analysis}
In this section, we discuss the finite-size scaling methodology used near the critical point to estimate the location of the transition and the universal critical exponents. 
For the chaotic-to-controlled phase transition, the magnetization $M_z$ changes from 0 in the chaotic phase ($p<p_c$) to +1 in the controlled phase ($p>p_c$), induced by the control rate $p$ at $p_c=0.5$. In the thermodynamic limit,  the correlation length $\xi\sim \abs{p-p_c}^{-\nu}$ and correlation time  $\xi_t\sim \abs{p-p_c}^{-\nu z}$ diverge at the transition in a universal fashion, where $\nu$ and $z$ are the correlation length  and dynamic exponents, respectively.
In the finite-size system $L$, the ``infrared'' cut-off introduces a length scale that is at most the system size $L$, leading to a modified scaling behavior near the critical point in space and time as $L\sim \abs{p-p_c}^{-\nu}$ and $t\sim L^z$, respectively. 
Therefore, the magnetization can be expressed in the form of $\overline{\expval{M_z}}\sim f_M(\abs{p-p_c}L^{1/\nu})$, as shown in \cref{fig:fig2}{\bf a}.
We choose a fitting range of $p\in[0.35,0.65] $ and $L\in[10,100]$ for experimental data, 
$p\in[0.4,0.6] $ and $L\in[10,40]$ for MPS calculations, and $p\in[0.4,0.6] $ and $L\in[10,100]$ for the theoretical first moment stat mech model.

For observables with a finite scaling dimension, e.g., that vanish at the critical point like $\mathcal{O}\sim \abs{p-p_c}^\beta$, we apply the finite-size scaling form $\mathcal{O}\sim L^{-\beta/\nu}f_\mathcal{O}(\abs{p-p_c}L^{1/\nu})$, where $\beta$ is the critical exponent corresponding to twice the scaling dimension of the operator under consideration in $\mathcal{O}$.
In Fig.~\ref{fig:fig2}{\bf b}, we choose a fitting range of $p\in[0.4,0.6] $ and $L\in[10,100]$ for experimental data, $p\in[0.4,0.6] $ and $L\in[10,40]$ for MPS calculations, and $p\in[0.4,0.6] $ and $L\in[10,100]$ for the theoretical first moment stat mech model.

For the dynamical data in \cref{fig:fig3}, we can compute the dynamical exponent $z$ (discussed above) that relates scaling of space and time through $t\sim L^z$.
Therefore, at the critical point $p=0.5$, we expect the growth of the magnetization follows $\overline{\expval{M_z(t)}} \sim f_M'(t/L^z)$ and with finite scaling dimension, we have $\widetilde{\mathrm{var}}_{Q}[{M_z}] \sim L^{-2\beta/\nu} f_Q'(t/L^z)$.
We choose a fitting range of $t\in[1,0.6L]$ as the early time dynamics to extract this growth before any finite-size effects appear in the data.

For more details on the numerical fitting procedure, we refer the reader to Sec.~\ref{sec:fss} in the Supplemental Material. 

\textbf{Matrix product state simulation}
We initialize an MPS as a product state in the open boundary condition to conceptually simulate a ``folded boundary condition''. Namely, the physical qubit index in an $L$-site MPS will change from $1,2, \dots, L$ to $1, L, 2, L-1, \dots, L/2-1, L/2$ (assuming even $L$).
The main advantage of this approach is that the efficient application of a two-qubit gate across the two ends, i.e.,  $1$ and $L$, to enforce periodic boundary conditions, with the trade-off being that all the nearest-neighbor gates are now at most next-nearest-neighbor gates.

We use the actual Haar random unitary obtained from QR decomposition in the MPS simulation. After each unitary gate, we set the truncation error as $10^{-10}$ (except for Fig.~\ref{fig:fig2}{\bf c} and Fig.~\ref{fig:fig3}{\bf c}), and let the bond dimension grow as needed.
In Fig.~\ref{fig:fig2}{\bf c} and Fig.~\ref{fig:fig3}{\bf c}, we set the truncation error as $10^{-15}$, and cap the bond dimension to $512$ to ensure the executability of the code even when it is inside the volume-law phase, and therefore, for $L\ge 20$, all bond dimensions are capped at $512$.
We use the \texttt{Julia} package \texttt{ITensors.jl} for the MPS simulations~\cite{itensor}.

Our simulation is performed on \texttt{AMD EPYC 9654}, which typically takes about 1.7s, 9.4s, 37s, and 103s to finish one single trajectory up to $2L^2$ steps for $L=10,20,30,$ and $40$ at $p=0.5$, respectively.
We numerically find complexity of the MPS simulation is $\mathcal{O}(L^{3.15})$. 

\textbf{Classical models}
We introduce several classical models to compare with the MPS simulation. 
In \cref{fig:fig2} and Fig.~\ref{fig:fig3}, we use the statistical mechanics (stat mech) model as a classical reference, as detailed in Sec.~\ref{sec:statmech} in the Supplemental Material.
In Fig.~\ref{fig:fig4}, we use the dephasing model as another classical reference, as detailed in Sec.~\ref{sec:dephasing} in the Supplemental Material. We give a brief overview of both classes of models here.

The stat mech model is useful for computing $ k$th moments of statistical quantities for integer $k$.
It uses the ``replica trick'' that relies on interchanging the expectation and trace:
\begin{align}
 \mathbb{E}_{C\sim \mathcal{B}(p)} \left[ \tr [(\rho_C \mathcal{O})^{\otimes k} ] \right] = \tr \left[ \mathbb{E}_{C\sim \mathcal{B}(p)} [\rho_C^{\otimes k}] \mathcal{O}^{\otimes k} \right],
\end{align}
where the left-hand side is precisely the $k$th moment of an observable $O$ and the right-hand side its $k$-copy version.
The model keeps track of the average $k$-copy state $\mathbb{E}_{C\sim \mathcal{B}(p)} [\rho_C^{\otimes k}]$ and can efficiently express through a classical random walk any update to this state.
In this work, we show how to simulate the average magnetization and the collision probability, which are first and second moment quantities, respectively, using this formalism (see Sec.~\ref{sec:statmech} for more details).

The dephasing model is the classical analog we use to benchmark results in \cref{fig:fig4}. 
It is, in a sense, the closest a classical computer can get to simulating the quantum dynamics. To make this precise: A quantum computer with full dephasing is equivalent to a classical Markov process on bitstrings.
Put simply, full dephasing sends $\rho \mapsto \diag(\rho)$ (only the diagonal of the density matrix survives).
$\diag(\rho)$, represented as a vector is a classical probability distribution on a bitstring $\bm{b}=\overline{b_1b_2\dots b_L}\equiv \sum_{i=1}^L b_i 2^{-i}$ is $p(b) = \braket{\bm{b}}{\diag(\rho)|\bm{b}}$.
Unitary evolution followed immediately by dephasing leads to $U\rho U^\dagger \mapsto \sum_{b'} T_U(b,b')p(b')$ where $T_U$ represents the Markov process associated with $U$, and resetting bit $b_L$ becomes simply $p(b_1\cdots b_L) \mapsto [p(b_1\cdots b_{L-1} 0) + p(b_1\cdots b_{L-1}1)] \delta_{b_L,0}$.

\textbf{Kullback-Leibler Divergence}
In Fig.~\ref{fig:fig4}, we compute the Kullback-Leibler (KL) divergence between two probability distributions $p(x)$ and $q(x)$.
Here, inside the controlled phase for the theoretical model, many samples lie exactly at $0$, and therefore, we can model the distribution as a mixture of a Dirac delta function at $0$ and a continuous distribution:
$p(x)=w_{p}\,\delta(x)+(1 - w_{p})\,p_{\text{cont}}(x)$,
and 
$q(x)=w_{q}\,\delta(x)+(1 - w_{q})\,q_{\text{cont}}(x)$,
where $p_{\text{cont}}(x)$ and $q_{\text{cont}}(x)$ are the continuous parts of the distributions.
Therefore, the KL divergence can be expressed as
\begin{equation}
    \begin{split}
        \mathcal{D}_{\text{KL}}(p, q)&=w_{p}\log(\frac{w_{p}}{w_{q}})
        +(1 - w_{p})\log(\frac{1 - w_{p}}{1 - w_{q}})\\
        &+(1 - w_{p})\mathcal{D}_{\text{KL}}(p_{\text{cont}},q_{\text{cont}}),
    \end{split}
\end{equation}
where the KL divergence of the continuous part is defined as 
\begin{equation}
    \mathcal{D}_{\text{KL}}(p_{\text{cont}} ,  q_{\text{cont}}) = \int p_{\text{cont}}(x) \log \frac{p_{\text{cont}}(x)}{q_{\text{cont}}(x)} dx,
\end{equation} 
Here, to estimate the continuous part of the probability density function $p_{\text{cont}}(x)$ and $q_{\text{cont}}(x)$, we use the kernel density estimation, and finally, using the bootstrapping method to estimate the error bar of the mean of the KL divergence.

\vspace{3mm}
\noindent\textbf{\large{}Acknowledgements}
\vspace{1mm}
\\
We thank Miles Stoudenmire for useful discussions and collaborations on related work, as well as Lucy Reading-Ikkanda of the Simons Foundation for the help with Fig.~1's design and execution. We also thank Oles Shtanko for helpful discussions.
This work was partially supported by the US-ONR grant No.~N00014-23-1-2357 (H.P.~and J.H.P.), the Army Research Office Grant No.~W911NF-23-1-0144 (K.A.~and J.H.P.), NSF Grant No.~DMR-2315063 (S.G.), and NSF CAREER Grants No.~DMR-2143635 (T.I.), and~DMR-2238895 (J.H.W.).
We are grateful to the DoD High Performance Computing Modernization Program (HPCMP) for providing us access to the Carpenter cluster to run all matrix product state and dephasing model simulations.
This work was performed in part at the Aspen Center for Physics, which is supported by National Science Foundation grant PHY-2210452 (T.I., J.H.W., J.H.P.) and at the Kavli Institute for Theoretical Physics (KITP), which is supported by grant NSF PHY-2309135 (J.H.P.). 
\bibliography{ref}

\clearpage
\begin{widetext}
  \begin{center}
    \Large Supplemental Material for ``\titletext''
  \end{center}
\end{widetext}

\section{Device Specifications}\label{sec:device_specifications}
\begin{figure*}[!ht]
    \vspace{4.6cm}
    \centering
    \includegraphics[width=\textwidth]{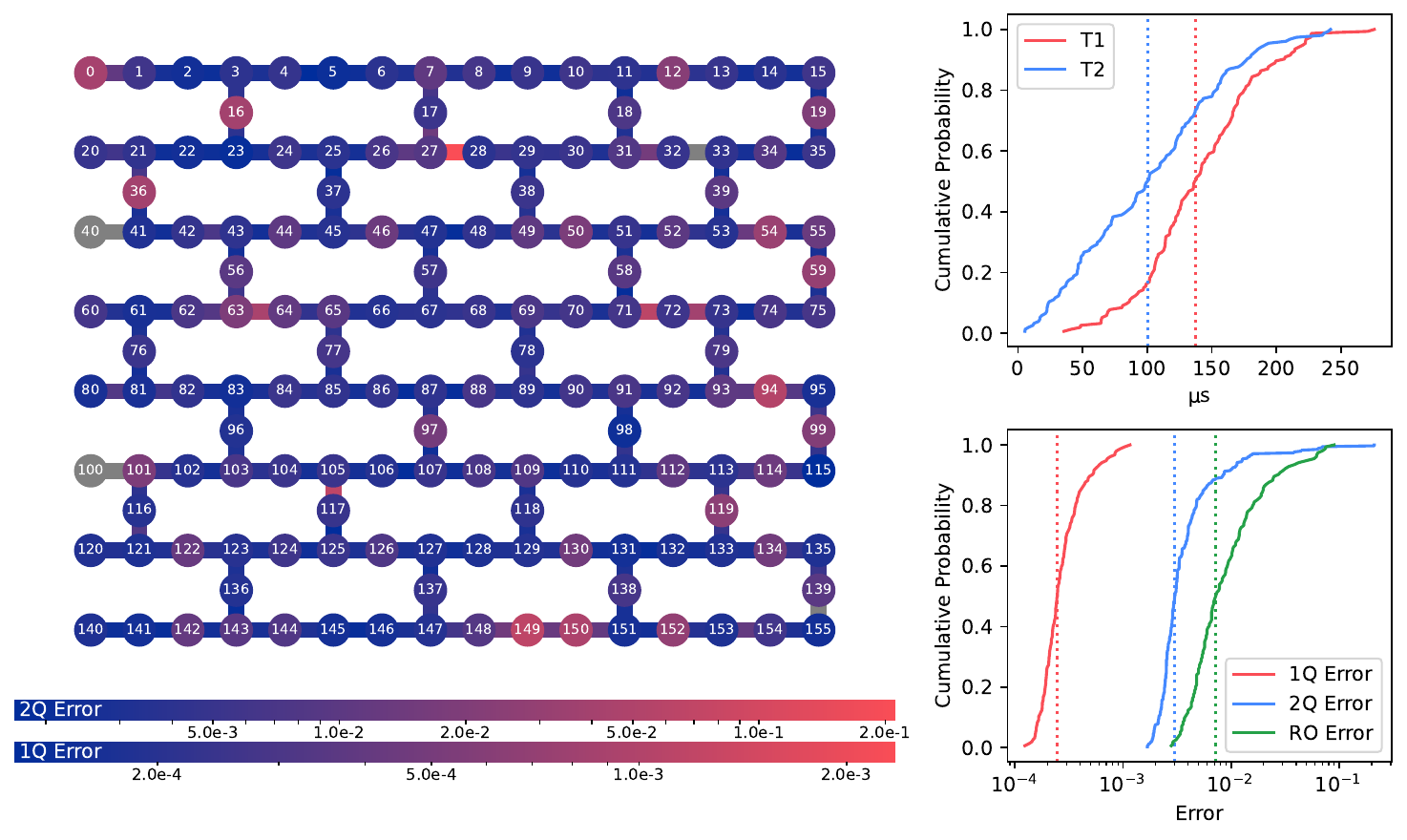}
    \caption{
    \textbf{Experimental device.}
    Device layout showing the chosen qubits and their specifications all the experimental realizations on $\texttt{ibm\_fez}$ corresponding to the data shown in Figure 2. Unused and un-characterized qubits and couplings and the couplings have been grayed out. 
    The mean $T_1$ and $T_2$ times, shown as dotted vertical lines in the plot above, were $1.4 \times 10^{2} \mu s$ and $1.0 \times 10^{2} \mu s$  respectively. Single and two-qubit durations were set at 24 ns and 84 ns and the readout duration was $1.6 \times 10^{3}$ ns. On average, single-qubit, two-qubit and readout errors were respectively $3.0 \times 10^{-4}$, $1.8 \times 10^{-2}$, $1.3 \times 10^{-2}$.
    }  \label{fig:connectivity}
    
\end{figure*}

The device connectivity, which represents a heavy-hexagonal lattice on 156 qubits, is shown in \cref{fig:connectivity}. The specifications for the device for the data collection days and the chosen qubits are also detailed in \cref{fig:connectivity}.

\section{Experimental Details}\label{sec:experimental_details}

\begin{figure}
    \centering
    \includegraphics[width=\linewidth]{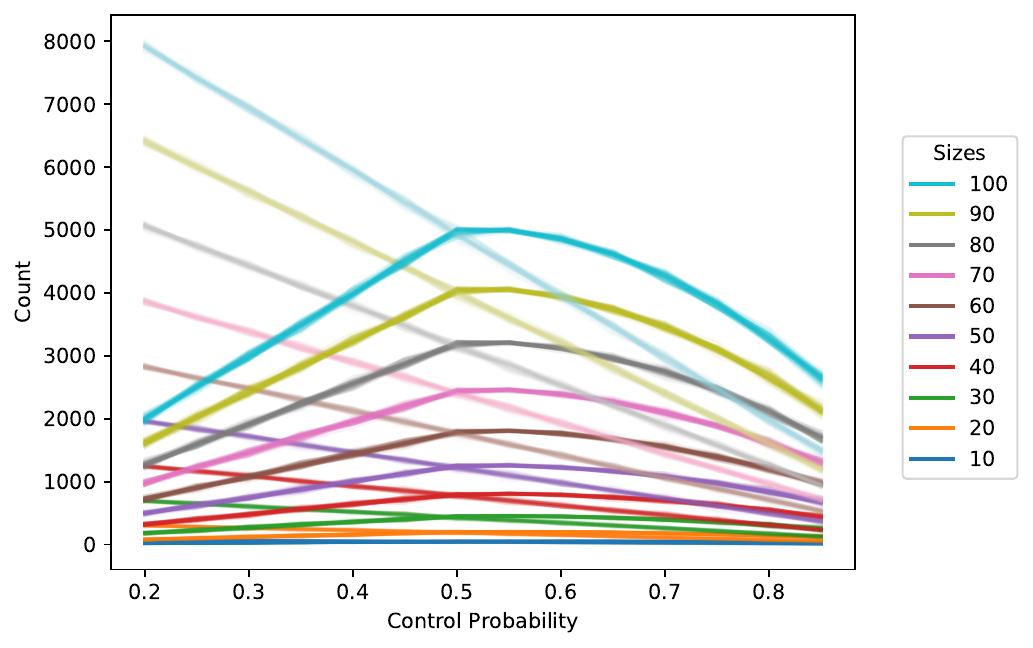}
    \includegraphics[width=\linewidth]{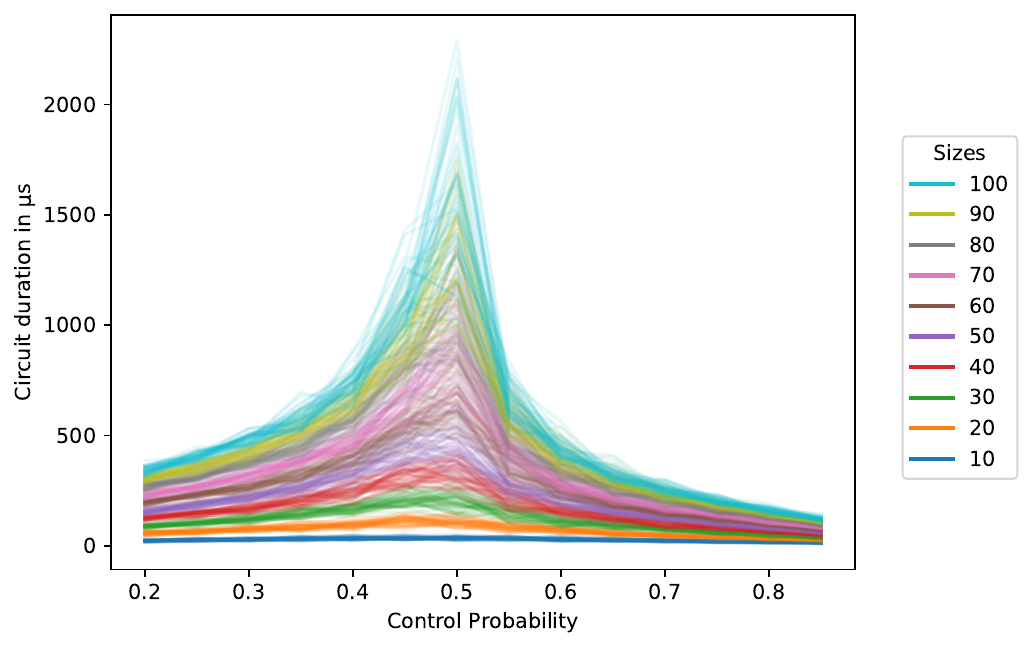}
    \caption{Top: the total number of entangling two-qubit operations (lighter curves) and measurement-induced conditional resets (darker curves). Bottom: Estimated circuit duration as a function of control probability for all 50 circuit realizations.}
    \label{fig:reset_cz_vs_L}
\end{figure}

At each problem size and control probability parameter, 50 different circuit realizations were constructed. Each circuit realization was repeated for 1000 shots. \cref{fig:reset_cz_vs_L} shows the average number of entangling
two-qubit operations and measurement-induced conditional resets at various values of the control parameter $p$. In particular, we considered problem sizes $L$ from 10 to 100 with a step size of 10 and $p$ from 0.2 to 0.85 with a step size of 0.05. The total number of time steps was set to $L^2$, and correspondingly, the two-qubit gate and reset count increased with $L$. Lower values of the control parameter require denser circuits with more entangling operations, with the maximum number of two-qubit gates reached at approximately 8000 gates at $p = 0.2$ for $ L = 100$. The maximum number of resets is reached at $p = 0.5$, with $L = 100$, using nearly 5000 resets. At $p_{\text{c}} = 0.5$, the circuit is most resource-intensive, with 5000 resets intermixed with 5000 two-qubit operations. Without transpilation, the number of resets should increase linearly with $p$, similar to the two-qubit operations that decrease linearly with $p$. However, a reset operation followed by another one is trivial, so we merge such consecutive resets, shown as the diminishing trend in the number of resets beyond $p = 0.5$ in \cref{fig:reset_cz_vs_L}. \cref{fig:reset_cz_vs_L} shows that the circuit duration is maximal at $p=0.5$ as the duration of reset is substantially longer compared to the two-qubit gate duration.

\section{Finite Size Scaling Analysis}\label{sec:fss}
With a given finite-size scaling ansatz, we can numerically fit and extract the critical exponents, such that all curves with different system sizes collapse onto a single universal curve after proper rescaling (see insets in Fig.~\ref{fig:fig2} and \ref{fig:fig3}).
For example, given a set of data points (observations) $\{\mathcal{O}_i^{(j)}\}$ at a tuning parameter $p_i$ and system size $L_i$, where $i=\{1,2,\ldots,N\}$ is the number of different parameters, and $j=\{1,2,\ldots,M\}$ is the number of different shots, the generic scaling form is 
\begin{equation}\label{eq:scaling}
    \overline{\mathcal{O}_i}\sim L_i^{-\beta/\nu} f[(p_i-p_c)L_i^{1/\nu}] ,
\end{equation} where $\overline{\mathcal{O}_i}\equiv \frac{1}{M} \sum_{j=1}^{M} \mathcal{O}_i^{(j)} $.
In order to find the optimal set of parameters $(\beta, \nu, p_c)$, we define the loss function as its reduced $\chi_\nu^2$ value to quantify the goodness of the fit, i.e., 
\begin{equation}\label{eq:chi2}
    \chi_\nu^2 = \frac{1}{N} \sum_{\iota=1}^{N} \left( \frac{\widetilde{\overline{\mathcal{O}}}_\iota - \widetilde{\mathfrak{O}}_\iota}{\widetilde{\sigma}_\iota} \right)^2,
\end{equation}  
where $\iota$ is the permutation of $i$ to ensure the ascending order of $x_{\iota} = (p_\iota-p_c)L_\iota^{1/\nu}$, and $\widetilde{\overline{\mathcal{O}}}_\iota = L_\iota^{\beta/\nu} \overline{\mathcal{O}}_\iota$.
$\widetilde{\mathfrak{O}}_\iota$ is the linear interpolation between the nearest two points $(x_{\iota-1}, \widetilde{\mathcal{O}}_{\iota-1} )$ and $(x_{\iota+1}, \widetilde{\mathcal{O}}_{\iota+1} )$ on both sides of $(x_\iota, \widetilde{\mathcal{O}}_\iota)$, namely, 
\begin{equation}
    \widetilde{\mathfrak{O}}_\iota = \frac{x_{\iota+1} - x_{\iota}}{x_{\iota+1}-x_{\iota-1}} \widetilde{\overline{\mathcal{O}}}_{\iota-1} + \frac{x_{\iota} -x_{\iota-1} }{x_{\iota+1}-x_{\iota-1}} \widetilde{\overline{\mathcal{O}}}_{\iota+1},
\end{equation} 
If $(x_\iota, \widetilde{\mathcal{O}}_\iota)$ is on the boundary, (i.e., $\iota = 1$ or $N$), we simply use single side linear interpolation.

The denominator in Eq.~\eqref{eq:chi2} $\tilde{\sigma}_\iota$ is the corresponding standard deviation of the interpolated $\widetilde{\mathfrak{O}}_\iota$, defined as
\begin{equation}
    \widetilde{\mathfrak{O}}_\iota^2 = \sigma_\iota^2+ \left( \frac{x_{\iota+1} - x_{\iota}}{x_{\iota+1}-x_{\iota-1}} \sigma_{\iota-1} \right)^2 + \left( \frac{x_{\iota} -x_{\iota-1} }{x_{\iota+1}-x_{\iota-1}}\sigma_{\iota+1} \right)^2,
\end{equation}
where $\sigma_\iota^2= \frac{1}{M}L^{-\beta/\nu}\sum_{j=1}^{M} \left( \mathcal{O}_\iota^{(j)} - \overline{\mathcal{O}}_\iota \right)^2$.

We minimize the loss function in Eq.~\eqref{eq:chi2} using the Levenberg-Marquardt algorithm~\cite{levenberg1944method,marquardt1963algorithm} provided by the Python package \texttt{lmfit}~\cite{newville2014lmfit} to obtain the optimal value of $\beta, \nu, p_c$ along with their error bars.

For the temporal scaling in Fig.~\ref{fig:fig3}(a,b), we can use the same procedure to collapse Eq.~\eqref{eq:scaling} but for the scaling function $\sim f(t/L^z)$ (and its generalization to finite scaling dimension). The corresponding error bar can thus be estimated by $\delta z = \nu^{-2}  \delta\nu $.

\section{Noise analysis for the magnetization}
In this section, we discuss the noise effect on the steady state of the magnetization, as shown in Fig.~\ref{fig:fig2}{\bf a}.
We model noise as a single-qubit depolarizing error channel $\Phi^{(1)}$ and two-qubit depolarizing error channel $\Phi^{(2)}$, i.e, 
\begin{equation}
    \Phi^{(n)}(\rho) = \left( 1-p_{e}^{(n)} \right)\rho +  \frac{p_{e}^{(n)}}{2^n} \mathds{1}_{2^n},
\end{equation}
acting on the single-qubit rotation $\hat{R}_\alpha(\theta)$ (see Eq.~\eqref{eq:approx_Haar}) and CZ gate in the unitary, respectively (see Eq.~\eqref{eq:B})
where $\rho $ is the density matrix for a single qubit, $\mathds{1}$ is the maximally mixed state up to a normalization factor, and $p_{e}^{(1)}$ ($p_{e}^{(2)}$) is the single-qubit (two-qubit) depolarizing error probability.
Using a Choi--Jamio\l{}kowski isomorphism, the density matrix $\rho=\sum_{i,j}\rho_{ij}\ketbra{i}{j}$ can be encoded as a ket state in in the double Hilbert space as $\kett{\rho}=\sum_{ij}\rho_{ij}\ket{i}\ket{j}$, where the quantum channel ${\Phi}^{(n)}$ is then described as 
\begin{equation}\label{eq:noise_Phi}
\hat{\Phi}^{(n)}\kett{\rho}\equiv\kett{{\Phi}^{(n)}(\rho)} =  \left[ \left( 1-p_{e}^{(n)} \right) + p_{e}^{(n)}\kettbbra{\mathds{1}}{\mathds{1}} \right]\kett{\rho},
\end{equation}
where $\kett{\mathds{1}}=2^{-n}\sum_{i=0}^{2^n-1}\kett{i}$ is the maximally entangled Einstein--Podolsky--Rosen (EPR) pair between the bra and ket space, and $n=1,2$ for the single and two-qubit channels, respectively.

Therefore, the two-qubit density matrix, where the corresponding Bernoulli map acts on, averaged over Haar randomness is
\begin{equation}\label{eq:noise_U}
    \E{\text{Haar}}{\hat{U}\kett{\rho}} = \E{\text{Haar}}{\hat{U}}\kett{\rho} = \mathbf{J}_{4} \oplus \mathbf{0}_{12} \kett{\rho}
\end{equation}
where $\hat{U}$ is the Haar random unitary $U$ in the double Hilbert space. Here, $\mathbf{J}_{4}$ is an all-one matrix of the size 4 by 4 acting on the basis in $\kett{\rho}$ that corresponds diagonal elements of $\rho$, and $\mathbf{0}_{12}$ is a zero matrix of the size 12 by 12 acting on those basis $\kett{\rho}$ associated with the off-diagonal elements of $\rho$ because the off-diagonal element of $\rho$ will be averaged to zero within Haar randomness.

Similarly, the single-qubit density matrix, on which the control map acts, is 
\begin{equation}\label{eq:noise_C}
    \hat{C}\kett{\rho} = (\kettbbra{00}{00} + \kettbbra{00}{11})\kett{\rho},
\end{equation}
where $\hat{C}$ is the control map $C$ in the double Hilbert space, such that it flips $\ketbra{1}$ to $\ketbra{0}$ while remain $\ketbra{0}$ intact.
 
We only track the diagonal elements of the density matrix because the off-diagonal elements will not mix with the diagonal ones.
Although the mapping is not strict in the classical model, however it suffices to study the change of order parameter in the control-induced phase transition.
We thus evolve the bit string probabilistically following the transition matrix in Eq.~\eqref{eq:noise_U} and~\eqref{eq:noise_C} with $1-p$ and $p$, in analogy to a Markov process.

\begin{figure*}[ht]
    \centering
    \includegraphics[width=6.8in]{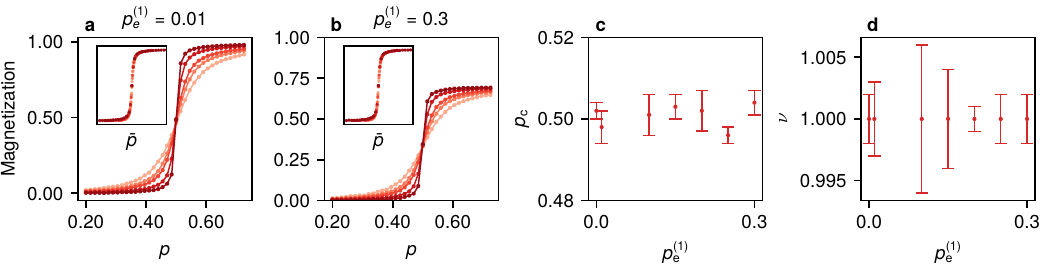}
    \caption{ 
        \textbf{Control Transition in the Presence of Noise}. 
        Results are shown for mechanics model calculations (red circles; $L=16-128$ ). 
        \textbf{a} The magnetization density averaged over initial states, measurement outcomes and circuits (denoted $\overline{\langle M_{z}\rangle}$) as a function of the control probability $p$ for various system sizes $L$, and single qubit depolarizing error $p_{e}^{(1)}=0.01$ and two qubit depolarizing error $p_{e}^{(2)}=0.001$. (Inset) Data collapse as a function of $\tilde{p}=L^{1/\nu}(p-p_c)$ with $p_c=0.498(1)$ and $\nu=1.000(2)$    
        \textbf{b} The magnetization density for single qubit depolarizing error $p_{e}^{(1)}=0.3$ and two qubit depolarizing error $p_{e}^{(2)}=0.03$.(Inset) Data collapse as a function of $\tilde{p}=L^{1/\nu}(p-p_c)$ with $p_c=0.504(3)$ and $\nu=1.000(2)$
        \textbf{c} The fitted $p_{c}$ as a function of the single qubit depolarizing error rate $p_{e}^{(1)}$, where $p_{e}^{(2)}=p_{e}^{(1)}/10$ is fixed. 
        \textbf{d} The fitted $\nu$ as a function of the single qubit depolarizing error rate $p_{e}^{(1)}.$
        }
    \label{fig:OP_noise}
\end{figure*}

We apply a single-qubit depolarizing noise channel in Eq.~\eqref{eq:noise_Phi} after each measurement with $p_e^{(1)}$ on the qubit where it was acted on, and two-qubit noise channel in Eq.~\eqref{eq:noise_U} after the Haar random gate with $p_e^{(2)}$.
Finally, we compute the magnetization averaged over different trajectories as shown in Figs. \ref{fig:OP_noise} \textbf{a} and \textbf{b} with the insets showing the data collapse.
We find that although the ferromagnetic fixed point state deep inside the controlled phase is suppressed, it still demonstrates a sharp phase transition near $p=0.5$, leading to the similar critical exponent, which is consistent with the experimental findings in Fig.~\ref{fig:fig2}{\bf a}.
In Figs. \ref{fig:OP_noise} \textbf{c} and \textbf{d}, we show fitted $p_c$ and $\nu$ as a function of the depolarizing error with a fixed relation $p_e^{(1)} = 10 p_e^{(2)}$.
We find that the depolarizing error does not affect the critical exponents within the error bar.
Experimentally, the error rates are reported as $10^{-4}$ to $10^{-2}$ for the two-qubit gate, and $10^{-3}$ to $10^{-2}$ for the measurement error, which implies our noise analysis does not underestimate the impact of errors.

\section{Classical Models}\label{sec:statmech}
In this work, we have considered circuits where the sequence of scrambling unitaries sample the Bernouli map $B_{i,i+1}$ in each application of the chaotic map that is chosen randomly from the distribution $\mathcal{D}$.
Under the assumption that this distribution over unitaries $\mathcal{D}$ is invariant under conjugation by single-qubit unitaries from either a one-design or a two-design, we can compute up to first-moment or second-moment quantities, respectively, of observables linear in $\rho$.
Specifically, the $k$'th moment of an observable is 
\begin{align}
\mathbb{E}_{C\sim \mathcal{B}(p)} \left[ \tr [(\rho_C \mathcal{O})^{\otimes k} ]\right],
\end{align}
where $\rho_C$ is the output of the circuit when the sampled unitaries are $C=(U_1,\ldots U_m)$. The state $\rho_U$ is in general mixed because the circuit has nonunitary channels corresponding to the control map in addition to the unitary channels coming from $U_1, \ldots U_m$.
By exchanging the trace and the expectation value, we can rewrite the $k$'th moment as
\begin{align}
 \tr \left[ \mathbb{E}_{C\sim \mathcal{B}(p)} [\rho_C^{\otimes k}] \mathcal{O}^{\otimes k} \right].
\end{align}
In this section, we deal with classical techniques for directly accessing the average $k$-copy state $\mathbb{E}_{C\sim \mathcal{B}(p)} [\rho_C^{\otimes k}]$ for $k=1$ and $2$, which lets us compute the first and second moments of the observable $\mathcal{O}$.
We denote the average $k$-copy state by $\bar{\rho}$.
Also denote the partial trace of $\bar{\rho}$ at a subset $A$ by $\bar{\rho}_A := \tr_{A^c} \bar{\rho}$.
It should be noted that this technique only gives us access to averages over unitaries from some suitable ensemble, and not the specific $k$'th moment for any particular sequence of unitaries $C$.

\subsection{First moment}
When $k=1$, it suffices to consider the average state after the chaotic and control maps.
Contained in the average $\mathbb{E}_{C\sim \mathcal{B}(p)}$ over the Bernoulli circuits lies an average $\mathbb{E}_{U_i \sim \mathcal{D}}$ over random circuits, which we first tackle.
After a chaotic map at sites $j,j'$, the 1-copy average state evolves as:
\begin{align}
 \bar{\rho}_{jj'} \to & \mathbb{E}_{U_i \sim \mathcal{D}}[U_i \bar{\rho}_{jj'} U_i^\dag] \\
 = & \frac{\mathds{1}_{jj'}}{4}.
\end{align}
This map is operationalised by replacing the classical representation of the state at these sites by a uniformly random bit.
On the other hand, after a control map on site $j$, the state $\bar{\rho}_j$ gets reset to $\ketbra{0}_j$, which is also easy to track using a classical bit.
These two simple rules suffice to yield a classical random walk that reproduces the order parameter $\mathbb{E}_{C\sim \mathcal{B}(p)}[\tr \rho_C \mathcal{O}]$ as well as the features of the shot noise seen in \cref{fig:fig2,fig:fig3}.

\begin{figure}
\centering
\includegraphics[width=\linewidth]{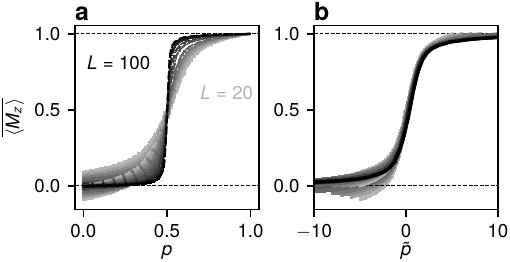}
\caption{
    \textbf{Magnetization in the first-moment stat mech model}. 
    \textbf{a} Magnetization density as a function of $p$ for different system sizes. ($L=20-100$ from light to dark curves.)
    \textbf{b} Finite size scaling of \textbf{a}. The fitted critical exponents are $p_c=0.4970(1)$ and $\nu=1.0000(1)$ with a fitting range of $p\in [0.4,0.6]$.    
} 
\label{fig:classicalmodel_firstmoment_orderparameter}
\end{figure}

We reproduce the magnetization density using the first-moment stat mech model as shown in Fig.~\ref{fig:classicalmodel_firstmoment_orderparameter}{\bf a} similar to Fig.~\ref{fig:fig2}{\bf a}.
In Fig.~\ref{fig:classicalmodel_firstmoment_orderparameter}{\bf b}, we perform the finite-size scaling analysis, which manifests the same critical behavior as in the inset of Fig.~\ref{fig:fig2}{\bf a}.

\begin{figure*}
\centering
\includegraphics[width=\linewidth]{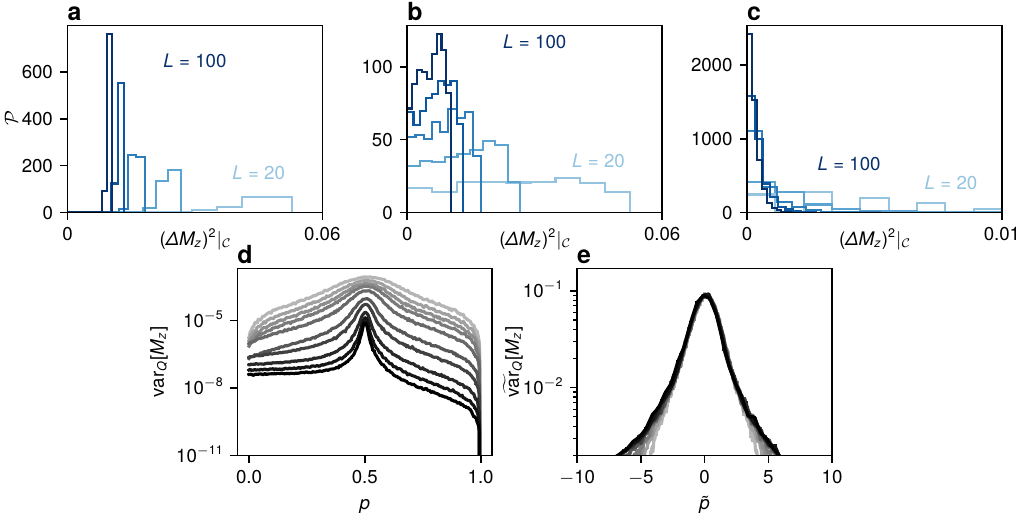}
\caption{
\textbf{Shot-to-shot fluctuations in the first-moment stat mech model.}
Behavior of shot-to-shot fluctuations of the order parameter from the classical stat-mech model for the first moment.
The fitted critical exponent is $p_c=0.5010(1)$, $\nu=1.00(2)$, and $\beta=1.011(4)$ with a fitting range of $p\in[0.4,0.6]$.   
}
\label{fig:classicalmodel_firstmoment_shotnoise}
\end{figure*}

{In Fig.~\ref{fig:classicalmodel_firstmoment_shotnoise}(a-c), we reproduce the distribution of the quantum fluctuations (shot-to-shot fluctuation) for the chaotic regime (Fig.~\ref{fig:classicalmodel_firstmoment_shotnoise}{\bf a}), the critical point (Fig.~\ref{fig:classicalmodel_firstmoment_shotnoise}{\bf b}), and the controlled regime (Fig.~\ref{fig:classicalmodel_firstmoment_shotnoise}{\bf c}), under various system sizes from $L=20$ to 100.
We find that, although they do not quantitatively reproduce the distribution of the shot-to-shot fluctuation (as we will show in Sec.~\ref{sec:dephasing} that the dephasing model is the closest classical model to the quantum case, yet the dephasing model still fails to capture all the features deep inside chaotic regime and controlled regime)---because it is by construction only accurate up to the first-moment while the shot-to-shot fluctuation is a second-moment quantity, it does capture the distribution near the critical points, and reproduces the same scaling dimension $\beta=1$, as shown in Fig.~\ref{fig:classicalmodel_firstmoment_shotnoise}(d,e).}

\subsection{Second moment}
We can also generalise the previous calculation to the second moment.
For both unitary \cite{nahum2017quantumentanglementgrowthrandom,nahum2018operatorspreadingrandomunitary,hunter-jones2019unitarydesignsstatisticalmechanics,dalzell2022randomquantumcircuitsanticoncentrate} and nonunitary random circuits \cite{dalzell2021randomquantumcircuitstransform,deshpande2022tightboundsconvergencenoisy,ware2023sharpphasetransitionlinear}, there is prior work computing several second-moment quantities such as expected purity, fidelity with a target state, the cross-entropy benchmark, and the variance of local observables.
In these settings, the calculation is greatly simplified by virtue of some assumptions on the ensemble of random circuits, which do not hold in our case.
We show how to overcome this hurdle by relaxing two fundamental assumptions underlying these calculations.

First, one major assumption in several of the above works is that each operation in the random circuit is chosen independently of the others.
In our setting, this assumption is invalid because after the application of every map, the choice of the next qubit(s) to apply the next map to depends on whether we applied the chaotic map or the control map.
In other words, the random circuit has important spatial correlations.
In fact, without this feature, our class of circuits would not feature a phase transition for a local observable.
We have already accounted for this correlation in the previous subsection when we calculate the first moment.
The tradeoff is that we do not analytically compute expectation values of first- or second-moment quantities but instead supply an \textit{efficient algorithm} to compute them.

Second, and more importantly, another assumption belying all above works is that the ensemble of random circuits is \textit{locally scrambling} \cite{angrisani2025classically}, i.e., is invariant under the insertion of random single-qubit gates from a 2-design after every operation.
In our case, this assumption is manifestly invalid, since after the control map there is no such single-qubit operation.
Indeed, if our ensemble of circuits were locally scrambling, there would be no signal to discern in a local observable and we would have $\tr \left[ \mathbb{E}_{C} [\rho_C] \mathcal{O} \right] = 0$.
We can nevertheless define a custom and novel stat-mech model for this situation, which can again be classically efficiently simulated.
We illustrate the technique by calculating the collision probability.

First, observe that it is enough to understand the physics of the two maps and the effect they have on the two-copy average state $\mathbb{E}_{C\sim \mathcal{B}(p)} [\rho_C^{\otimes 2}] =: \bar{\rho} $.
Since we are not considering any postselection and hence probe quantities linear in $\rho$, the effect of these maps may be computed by considering an input reduced density matrix supported at the location of these maps.
For example, for a control map acting on qubit $j$, it is enough to consider its action on $\bar{\rho}_j$, which we recall is defined as $\bar{\rho}_j = \tr_{j^c} \bar{\rho} $.
The collision probability is given by 
\begin{align}
 \sum_{\bm x} p_{\bm x}^2 = \sum_{\bm x} \tr \rho^{\otimes 2} \ketbra{\bm x^{\otimes 2}} \nonumber \\
 \implies \mathbb{E}_{\mathcal{C}} \sum_{\bm x} p_{\bm x}^2 = \sum_{\bm x} \tr \bar \rho \ketbra{\bm x^{\otimes 2}}.
\end{align}
A certain subset $A$ of the $[n]$ qubits have a chaotic map last applied to them, and their complement $A^c = [n] \backslash A $ have a control map last applied to them.
Since there is a control map on this subset $A^c$, these qubits are in the state $\bar{\rho}_{A^c} =  \left( \ketbra{00\ldots 0}_{A^c} \right)^{\otimes 2}$.
The two-copy state on the other qubits in $A$ lies in $\mathrm{span}\{I,S\}^{\abs{A}}$, where $I$ and $S$ are the identical and SWAP.
Denote the entire two-copy state
\begin{align}
\bar{\rho} = \left( \sum_{\bm x \in \{0,1\}^{\abs{A}}} c_{\bm x} S^{\bm x} \right) \otimes \left( \ketbra{00\ldots 0}_{A^c} \right)^{\otimes 2}.
\end{align}
The expected collision probability is 
\begin{align}
 \sum_{\bm x: \bm{x}_{A^c} = 00\ldots 0, \bm y} \tr c_{\bm y}  S^{\bm y} \ketbra{\bm x^{\otimes 2}} = 2^{\abs{A}}\sum_{\bm y} c_{\bm y}. \label{eq_contribution_to_cp}
\end{align}

We have seen that it suffices to keep track of the quantity $\sum_{\bm y} c_{\bm y}$.
We compute the action of the two maps and state some useful facts about the dynamics of the two-copy state through the following claims. 

\begin{claim}[Two-copy state after chaotic map]
For any arbitrary subset $A \subseteq [n]$, the reduced two-copy state in the circuit dynamics after the application of a chaotic map lies in the space $\mathrm{span}\{I, S\}^{\otimes \abs{A}}$, where $\abs{A}$ is the number of qubits in $A$ and the operators $I$ and $S $ are the identity and SWAP operators, respectively, acting on a qubit and its copy.
\end{claim}
This follows from the property that the chaotic map is locally scrambling.
For a locally scrambling distribution over two-qubit unitaries, the distribution is invariant under insertion of random single-qubit gates from a 2-design.
Let us denote these $V$.
Averaging over these random single-qubit gates, we project on to the symmetric subspace, which is spanned by $\{I,S\}^n$. 
We can explicitly compute this projection using linearity of trace:
\begin{align}
 & \mathbb{E}_V[(V \otimes V) \bar{\rho} (V \otimes V)^\dag] \nonumber \\
 = & \frac{\tr  \bar{\rho} - \frac{1}{2} \tr \bar{\rho}S }{3} I + \frac{\tr  \bar{\rho} S - \frac{\tr  \bar{\rho}}{2}}{3} S.
\end{align}

\begin{claim}[Two-copy state after control map]
The reduced two-copy state in the circuit dynamics after the application of a control map on qubit $j$ is  $\ketbra{0}^{\otimes 2}$.
\end{claim}
This fact follows from the fact that the control map is a reset operation, producing the state
\begin{align}
\ketbra{00} = \frac{1}{4}(II + IZ + ZI + ZZ).
\end{align}

These two claims, together with the observation that at every point in time, the reduced two-copy state at a particular site is either $\ketbra{0}^{\otimes 2}$ or in $\{I,S\}$, suffice for the classical simulation of the collision probability.

When the chaotic map is a Haar-random two-qubit unitary, we have the following rules for the classical random walk on $\{I,S\}^{2}$:
\begin{align}
 II \to II ;  \quad SS \to SS, \nonumber \\
 IS, SI \to \frac{4}{5} \left( \frac{II +SS}{2}\right). \label{eq_statmech_update_chaoticmap}
\end{align}
This yields the following update rules for the collision probability. We examine three cases.
 i) Both qubits are in a pure state, ii) One qubit is pure and the other already lies in $\{I,S\}$, iii) Both qubits are in $\{I,S\}$.
For convenience, we note that if the initial state is pure, for example in the $\ket{0}$ state, upon applying a single-qubit Haar average we get:
\begin{align}
 \ketbra{0^{\otimes 2}} \to \frac{I+S}{6}.
\end{align}

i) In this case, the input two-copy average state at sites $j,j'$ where the chaotic map is applied is $\ketbra{0}^{\otimes 2}_{j} (aI +bS)_{j'}$.
In the input state, the contribution of these qubits to the collision probability is $1 \times 2(a+b)$. Here the factor 1 comes from the qubit $j$ being in a pure state, while the second factor $2(a+b)$ comes from the qubit $j'$ being in $(aI +bS)_{j'}$ (see \cref{eq_contribution_to_cp}). 
After application of the chaotic map, the output two-copy state is given by
\begin{align}
  II \left( \frac{7a}{30} + \frac{2b}{30} \right) + SS \left( \frac{2a}{30} + \frac{7b}{30} \right).
\end{align}
This yields a contribution to the collision probability of 
\begin{align}
 2^2 \times \left( \frac{9a}{30} + \frac{9b}{30} \right) = \frac{6}{5}(a+b).
\end{align}
Therefore, the contribution to the collision probability has decreased from $2 \to \frac{6}{5}$, a factor of $\frac{3}{5}$.

ii) The input two-copy average state is $\ketbra{00}_{jj'}^{\otimes 2}$ since both sites last had a control map applied to them. The initial contribution to the collision probability is $1$.
After the chaotic map, the output two-copy state is now
\begin{align}
\frac{II + SS}{20}.
\end{align}
This has a contribution of $2^2 \times \frac{2}{20} = \frac{2}{5}$ to the collision probability.
In this case, the reduction in collision probability is by a factor of $\frac{2}{5}$.

iii) If the reduced two-copy state on qubits $j,j'$ is in $\mathrm{span}\{I,S\}^{\otimes 2}$, we apply the rules in \cref{eq_statmech_update_chaoticmap}. The classical simulation of the stat-mech model samples ``words'' in $\{I,S\}^{\otimes 2}$. If the sampled words are $II$ (respectively, $SS$), the update rules keep them in $II$ (resp.\,$SS$). The collision probability is left unchanged.
If the word is either $IS$ or $SI$, we pick with probability $1/2$ each whether to update it to either $II$ or $SS$.
The factor $4/5$ from \cref{eq_statmech_update_chaoticmap} is accounted for by multiplying the collision probability by $4/5$ in this case.

Lastly, we examine what happens when we apply a control map at site $j$.
If the reduced two-copy state at site $j$ is already in $\ketbra{0}^{\otimes 2}$, the control map leaves it unchanged.
If we have $\bar{\rho}_j = \frac{a}{4}I+\frac{b}{2}S$ (with $\tr \bar{\rho}_j=  a + b =1$), we effectively sample $I$ with probability $a$ and $S$ with probability $b$.
When the sampled word is $I$, the contribution to the collision probability prior to the control map is $2\times \frac{a}{4} = \frac{a}{2}$, and the final contribution is $a$.
Similarly, when the sampled word is $S$, the prior contribution to the collision probability is $2\times \frac{b}{2} = b$, and the final contribution is $b$.
Therefore, upon application of the control map, if the input sampled word is $I$, we pick up a factor of 2 in the collision probability. If the sampled word is $S$, we leave the collision probability unchanged.

Thus, we can see that it is the control map that results in an increase in collision probability. This is natural, because the control map tends to reset qubits to $\ket{0}$, and in the extreme case the output distribution is peaked on bitstrings with many $\ket{0}$s.

\begin{figure}[ht]
    \centering
    \includegraphics[width=3.4in]{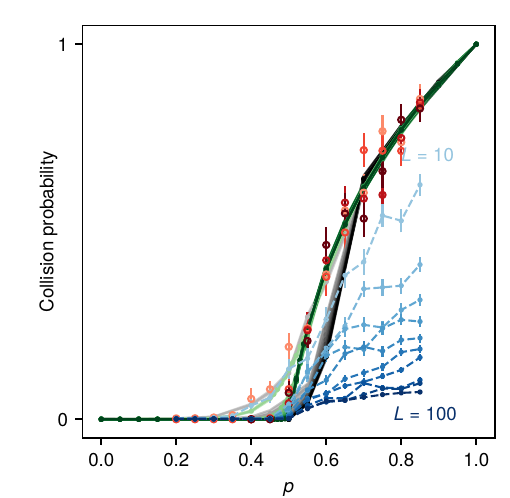}
    \caption{
    \textbf{Collision probability.}
    Blue dashed curves for the experimental data. Red curves for MPS simulation using the same 50 circuits. Green curves for the first moment stat mech model, and the gray curves for the second moment stat mech model.
    }
    \label{fig:CP_vs_p}
\end{figure}

These rules, taken together, allow for a classical simulation of the expected collision probability, as shown in \cref{fig:CP_vs_p}. At $p=0$, we numerically find that the predicted collision probability is close to the ideal value of $2/(2^n + 1) \approx 2^{-(n-1)}$, and at $p=1$, it approaches the value $1$, which is expected for the fixed point state $\ket{00 \ldots 0}$, which has a peaked distribution.
We also roughly see that in the control phase $p>0.5$, the collision probability is nonzero even for large $L$.
This illustrates that the classical bitstring distribution in this phase has a large peak, even though it may be formally supported on many other bitstrings. 
When $p<0.5$, the collision probability is exponentially small and the stat mech model needs a large number of samples to precisely predict the ideal expected collision probability.
This behavior is also expected, since in this phase the classical distribution is supported on exponentially many bitstrings.

Finally, we note that we can also extract similar physics from the first moment stat mech model, although we do not expect it to accurately model second moment quantities.
Formally, we predict a collision probability from the first moment stat mech model by pretending that the true state is given by $\ket{00\ldots 0}_{A}\otimes \prod_{j \notin A} (\frac{\mathds{1}_j}{2})$, where $A$ is again the set of qubits that last encountered a reset operation.
This predicts a collision probability of $2^{-(n-\abs{A})}$.
For $p=0$, since $\abs{A}=0$, this predicts a collision probability of $2^{-n}$, which differs from the ideal collision probability of $\approx 2^{-1-n}$.
This difference is not detectable from the experiment because at large system sizes ($L\gtrsim 40$), the number of outcomes is much larger than the number of shots taken.

\section{Dephasing model}\label{sec:dephasing}

In this section, we introduce the details of the noise model in the dephasing channel, which we dub the ``dephasing model'' for brevity. The motivation of the dephasing model is to try and find a noise channel that is as close as possible to classically ``spoofing'' the observer when trying to measure the quantum fluctuations defined in Eq.~\eqref{eq:quantum_fluctuation_Mz}. This is distinct from the statistical mechanics model we have investigated, which is more akin to the depolarizing noise channel that averages over Haar unitaries to define the classical process (see Sec.~\ref{sec:statmech}). Instead, we can retain more classical fluctuations by focusing on a ``classical'' circuit (i.e., without quantum coherence) while keeping the notation of ``circuit realization'' sharp (i.e., without first taking the Haar random average of the density matrix).

Therefore, the dephasing model only focuses on the evolution of the \textit{amplitude} of the wave function, which is an $L_1$-normed probability mass function $\ketp{\psi}$ supported on the computational basis states $\bm{b} = \overline{b_1b_2\dots b_L}\equiv \sum_{i=1}^L b_i 2^{-i} $.  For example, for a Haar random unitary, $\hat{U}\sim $U(4), the transition matrix $\hat{U_t}$ can be constructed as 
\begin{equation}\label{eq:U_t}
    [\hat{U_t} ]_{ij} = \abs{\hat{U}_{ij}}^2,
\end{equation} 
such that $(\bm{b}|\hat{U_t}|\psi ) \equiv \abs{\mel{\bm{b}}{\hat{U}}{\psi}}^2$.

For the control map, we simply reset the corresponding qubit $i$. Namely, 
\begin{equation}\label{eq:C}
    \hat{C} = \bigotimes_{j=1}^{L} \left[ {\begin{pmatrix}
     1 & 0 \\
     0 & 1 
    \end{pmatrix}(1-\delta_{i,j}) + \begin{pmatrix}
        1 & 1\\
        0 & 0\\
    \end{pmatrix}
     \delta_{i,j}}  \right],
\end{equation}
where the order of the basis on each site is 0 and 1.

In the end, we sample the same circuit by fixing the realization of $\hat{U}_T$ and its spacetime position.
The ``quantum fluctuations'' are in analogy defined as 
\begin{equation}
    (\Delta M_z)^2\vert_{\C} = \E{\psi}{M_z^2} - \left( \E{\psi}{M_z} \right)^2
\end{equation}
where $\E{\psi}{\mathcal{O}} = \sum_{\bm{b}}  p_{\bm{b}}(\psi) \mathcal{O}_{\bm{b}} $ , and $p_{\bm{b}}(\psi) = \left( \bm{b} | \psi\right)$.

In numerics, since Eq.~\eqref{eq:U_t} and Eq.~\eqref{eq:C} are both essentially Markov processes, we can simulate the time evolution of a single bit string $\bm{b}$ under the stochastic dynamics of transition matrix $\hat{U}_t$ and $\hat{C}$ in replacement of the full probability mass function simulation of $\ketp{\psi}$.

\section{System size dependence of the Kullback-Leibler divergence}\label{sec:KL}
\begin{figure*}[ht]
    \centering
    \includegraphics[width=6.8in]{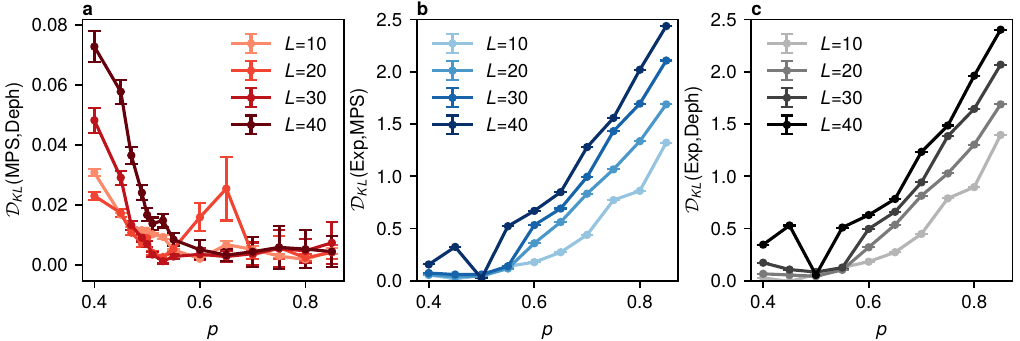}
    \caption{ 
        \textbf{The system size dependence of the Kullback-Leibler divergence}.
        \textbf{a} The KL divergence between the matrix product state (MPS) simulation and dephasing model (Deph) for various system sizes.
        \textbf{b} The KL divergence between the experimental (Exp) data and matrix product state (MPS) simulation and for various system sizes.
        \textbf{c} The KL divergence between the experimental (Exp) data and dephasing model (Deph) for various system sizes.
        }
    \label{fig:KL_L}
\end{figure*}

The Kullback-Leibler (KL) divergence is a measure of how one probability distribution differs from a second, reference probability distribution. 
In our case, we can use the KL divergence to quantify the difference among the probability density function of the variance of the magnetization obtained from the MPS simulation, the dephasing model, and the experimental data.
In Fig.~\ref{fig:KL_L}{\bf a}, we show the KL divergence between the MPS simulation and the dephasing model for various system sizes.
We find that the magnitude of the KL divergence increases with system size inside the chaotic phase, while remains zero within the error bar in the controlled phase, which implies that the dephasing model only captures the behavior of the system in the controlled phase, where it becomes purely classical dynamics, while the MPS simulation remains a true quantum model inside the chaotic phase which cannot be classically predicted.
In Fig.~\ref{fig:KL_L}{\bf b} and {\bf c}, we compare the ``experimental explanability'' of the quantum ``MPS'' simulation versus that of the classical dephasing model, and we find that the latter KL divergence is always larger than the former inside the chaotic phase for all system sizes.
In the controlled phase, both KL divergence increases to a large value because of the intrinsic noise in the experiments that will always prevent the full suppression of quantum fluctuations.

\end{document}